\documentclass{JHEP3}
\usepackage{epsfig}
\newcommand{\Li}[1]{\mathop{\mathrm{Li}}\nolimits_{#1}}
\newcommand{\mathph}[1]{\href{http://xxx.lanl.gov/abs/math-ph/#1}{\tt math-ph/#1}}
\preprint{SFB/CPP-11-32, TTP11-07}
\title{Simultaneous decoupling of bottom and charm quarks}
\author{Andrey G.\ Grozin$^{a,b}$,
Maik H\"oschele$^b$,
Jens Hoff$^b$
and Matthias Steinhauser$^b$\\
$^a$ Budker Institute of Nuclear Physics, Novosibirsk 630090, Russia\\
$^b$ Institut f\"ur Theoretische Teilchenphysik,
Karlsruher Institut f\"ur Technologie,\\
D-76128 Karlsruhe, Germany\\
E-mail: \email{A.G.Grozin@inp.nsk.su},
\email{hoeschele@particle.uni-karlsruhe.de},
\email{jens@particle.uni-karlsruhe.de}
and \email{matthias.steinhauser@kit.edu}}

\abstract{We compute the decoupling relations for 
  the strong coupling, the light quark masses, the gauge-fixing parameter,
  and the light fields in QCD with heavy charm and bottom quarks to three-loop  
  accuracy taking into account the exact dependence on $m_c/m_b$.
  The application of a low-energy theorem allows the extraction of the
  three-loop
  effective Higgs-gluon coupling valid for extensions of the Standard Model with
  additional heavy quarks from the decoupling constant of $\alpha_s$.
}


\begin{document}


\section{Introduction}
\label{S:Intro}

QCD where all six quark flavours are treated as active degrees of freedom is rarely
used in practical applications.
If the characteristic energy scale is below some heavy-flavour masses,
it is appropriate to construct a low-energy effective theory
without those heavy flavours.
The Lagrangian of this theory has the same form as the one of QCD
plus corrections suppressed by powers of heavy-quark masses.
Usually, heavy flavours are decoupled one at a time which results 
in a tower of effective theories,
each of them differ from the previous one by integrating out
a single heavy flavour.
The parameters of the Lagrangian of such an effective low-energy QCD
($\alpha_s(\mu)$, the gauge fixing parameter $a(\mu)$,
light-quark masses $m_i(\mu)$)
are related to the parameters of the underlying theory
(including the heavy flavour) by so-called decoupling relations.
The same holds for the light fields (gluon, ghost, light quarks)
which exist in both theories.
QCD decoupling constants are known at
two-~\cite{Bernreuther:1981sg,Larin:1994va,Chetyrkin:1997un}, 
three-~\cite{Chetyrkin:1997un}
and even four-loop order~\cite{Schroder:2005hy,Chetyrkin:2005ia}.

The conventional approach just described 
ignores power corrections in ratios of heavy-quark masses.
Let us, e.g., consider the relation
between $\alpha_s^{(3)}$ and $\alpha_s^{(5)}$ (the superscript denotes the
number of active flavours).
Starting from three loops, there are diagrams
containing both $b$- and $c$-quark loops
which depend on $m_c/m_b$.
The power correction $\sim(\alpha_s/\pi)^3\,(m_c/m_b)^2$
is not taken into account in the standard approach,
although, it might be comparable with the four-loop corrections of order
$(\alpha_s/\pi)^4$.
In the present paper, we consider $(m_c/m_b)^n$ power corrections
at three loops by decoupling $b$ and $c$ quarks in a single step.

Of course, the results presented in this paper are generic and apply to any
two flavours which are decoupled simultaneously from the QCD Lagrangian.
Our full theory is QCD with $n_l$ light flavours,
$n_c$ flavours with mass $m_c$, and $n_b$ flavours with mass $m_b$
(in the real world $n_c=n_b=1$). Furthermore we introduce the total number of
quarks $n_f=n_l+n_c+n_b$.
We study the relation of full QCD to the low-energy effective theory
containing neither $b$ nor $c$.

The bare gluon, ghost and light-quark fields in the effective theory
are related to the bare fields in the full theory by
\begin{equation}
  A_0^{(n_l)} = \left(\zeta_A^0\right)^{1/2} A_0^{(n_f)}\,,\quad
  c_0^{(n_l)} = \left(\zeta_c^0\right)^{1/2} c_0^{(n_f)}\,,\quad
  q_0^{(n_l)} = \left(\zeta_q^0\right)^{1/2} q_0^{(n_f)}\,,
  \label{Intro:fields0}
\end{equation}
where the bare decoupling constants are computed
in the full theory via~\cite{Chetyrkin:1997un} 
\begin{eqnarray}
  \zeta_A^0(\alpha_{s0}^{(n_f)},a_0^{(n_f)}) &=& 1 + \Pi_A(0)
    = \left[Z_A^{\rm os}\right]^{-1}\,,
  \nonumber\\
  \zeta_c^0(\alpha_{s0}^{(n_f)},a_0^{(n_f)}) &=& 1 + \Pi_c(0)
     = \left[Z_c^{\rm os}\right]^{-1}\,,
  \nonumber\\
  \zeta_q^0(\alpha_{s0}^{(n_f)},a_0^{(n_f)}) &=& 1 + \Sigma_V(0)
     = \left[Z_q^{\rm os}\right]^{-1}\,,
  \label{Intro:zetafields0}
\end{eqnarray}
with $\alpha_{s0}=g_0^2/(4\pi)^{1-\varepsilon}$;
$\Pi_A(q^2)$, $\Pi_c(q^2)$ and $\Sigma(q) = \rlap/q \Sigma_V(q^2)+m_{q0} \Sigma_S(q^2)$
are the (bare) gluon, ghost and light-quark self-energies
(we may set all light-quark masses to 0 in $\Sigma_V$ and $\Sigma_S$).
The fields renormalized in the on-shell scheme coincide in both theories;
therefore, the bare decoupling coefficients~(\ref{Intro:zetafields0})
are the ratio of the on-shell renormalization constants of the fields.
In the effective theory all the self-energies vanish at $q=0$
(they contain no scale), and the on-shell $Z$ factors are exactly 1.
In the full theory, only diagrams with at least one heavy-quark loop 
survive.\footnote{At low $q\ne0$, the self-energies in the full theory are given by sums
of contributions from various integration regions, see, e.\,g., \cite{Smirnov:2002pj};
the contribution we need comes from the completely hard region,
where all loop momenta are of order of heavy-quark masses.}

Next to the fields also the parameters of the full and effective QCD
Lagrangian are related by decoupling constants
\begin{equation}
  \alpha_{s0}^{(n_l)} = \zeta_{\alpha_s}^0 \alpha_{s0}^{(n_f)}\,,\quad
  a_0^{(n_l)} = \zeta_A^0 a_0^{(n_f)}\,,\quad
  m_{q0}^{(n_l)} = \zeta_m^0 m_{q0}^{(n_f)}\,,
\label{Intro:params0}
\end{equation}
where $a$ is the gauge parameter defined through the gluon propagator
\begin{equation}
  D_{\mu\nu}(k) = -\frac{i}{k^2}\, \left( g_{\mu\nu} - (1-a)\,
  \frac{k_\mu k_\nu}{k^2} \right)\,.
  \label{eq::gluon_prop}
\end{equation}
The bare decoupling constants in Eq.~(\ref{Intro:params0}) are computed with
the help of~\cite{Chetyrkin:1997un}
\begin{eqnarray}
  \zeta_{\alpha_s}^0(\alpha_{s0}^{(n_f)}) &=&
  \left(1+\Gamma_{A\bar{c}c}\right)^2 \left(Z_c^{\rm os}\right)^2 Z_A^{\rm os} =
  \left(1+\Gamma_{A\bar{q}q}\right)^2\left(Z_q^{\rm os}\right)^2 Z_A^{\rm os} =
  \left(1+\Gamma_{AAA}\right)^2
  \left(Z_A^{\rm os}\right)^3\,,
  \nonumber\\
  \zeta_m^0(\alpha_{s0}^{(n_f)}) &=& Z_q^{\rm os} \left[1 - \Sigma_S(0)\right]\,.
  \label{Intro:zetaparams0}
\end{eqnarray}
The $A\bar{c}c$, $A\bar{q}q$ and $AAA$ proper vertex functions
are expanded in their external momenta,
and only the leading non-vanishing terms are retained.
In the low-energy theory they get no loop corrections,
and are given by the tree-level vertices of dimension-4 operators in the Lagrangian.
In full QCD (with the heavy flavours) they have just one
colour and tensor (and Dirac) structure, namely, that of the tree-level vertices
(if this were not the case, the Lagrangian of the low-energy theory
would not have the usual QCD form\footnote{The $A\bar{q}q$ vertex at 0-th order
in its external momenta obviously has only the tree-level structure.
For the $A\bar{c}c$ vertex at the linear order in external momenta,
this statement is proven in Appendix~\ref{S:Ghost}.
The $AAA$ vertex at the linear order in its external momenta can have,
in addition to the tree-level structure, one more structure:
$d^{a_1 a_2 a_3}(g^{\mu_1 \mu_2}k_3^{\mu_3}+\mbox{cycle})$;
however, the Slavnov--Taylor identity
${\langle}T\{\partial^\mu A_\mu(x),\partial^\nu A_\nu(y),\partial^\lambda A_\lambda(z)\}{\rangle}=0$
leads to $\Gamma^{a_1 a_2 a_3}_{\mu_1 \mu_2 \mu_3} k_1^{\mu_1} k_2^{\mu_2} k_3^{\mu_3} = 0$
(see Ref.~\cite{Pascual:1984zb}), thus excluding this second structure.}).
Therefore, we have the tree-level vertices times
$(1+\Gamma_i)$, where loop corrections $\Gamma_i$ contain at least one
heavy-quark loop. 
The various versions in the first line of Eq.~(\ref{Intro:zetaparams0})
are obtained with the help of the QCD Ward identities involving three-particle
vertices. In our calculation we restrict ourselves for convenience to the
ghost--gluon vertex.
Note that the gauge parameter dependence cancels in $\zeta_{\alpha_s}^0$ and
$\zeta_m^0$ whereas the individual building blocks in
Eq.~(\ref{Intro:zetaparams0}) still 
depend on $a$. This serves as a check of our calculation.

The $\overline{\mbox{MS}}$ renormalized parameters and fields
in the two theories are related by
\begin{eqnarray}
\alpha_s^{(n_l)}(\mu') &=& \zeta_{\alpha_s}(\mu',\mu) \alpha_s^{(n_f)}(\mu)\,,\quad
a^{(n_l)}(\mu') = \zeta_A(\mu',\mu) a^{(n_f)}(\mu)\,,
\nonumber\\
m_q^{(n_l)}(\mu') &=& \zeta_m(\mu',\mu) m_q^{(n_f)}(\mu)\,,\quad
A^{(n_l)}(\mu') = \zeta_A^{1/2}(\mu',\mu) A^{(n_f)}(\mu)\,,
\nonumber\\
c^{(n_l)}(\mu') &=& \zeta_c^{1/2}(\mu',\mu) c^{(n_f)}(\mu)\,,\quad
q^{(n_l)}(\mu') = \zeta_q^{1/2}(\mu',\mu) q^{(n_f)}(\mu)\,,
\label{Intro:ren}
\end{eqnarray}
where we allow for two different renormalization scales in the full and
effective theory. The finite decoupling constants are obtained by
renormalizing the fields and parameters in Eqs.~(\ref{Intro:zetafields0})
and~(\ref{Intro:params0}) which leads to
\begin{eqnarray}
\zeta_{\alpha_s}(\mu',\mu) &=&
\left(\frac{\mu}{\mu'}\right)^{2\varepsilon}
\frac{Z_{\alpha}^{(n_f)}\left(\alpha_s^{(n_f)}(\mu)\right)}%
{Z_{\alpha}^{(n_l)}\left(\alpha_s^{(n_l)}(\mu')\right)}
\zeta_{\alpha_s}^0\left(\alpha_{s0}^{(n_f)}\right)\,,
\nonumber\\
\zeta_m(\mu',\mu) &=&
\frac{Z_m^{(n_f)}\left(\alpha_s^{(n_f)}(\mu)\right)}%
{Z_m^{(n_l)}\left(\alpha_s^{(n_l)}(\mu')\right)}
\zeta_m^0\left(\alpha_{s0}^{(n_f)}\right)\,,
\nonumber\\
\zeta_A(\mu',\mu) &=&
\frac{Z_A^{(n_f)}\left(\alpha_s^{(n_f)}(\mu),a^{(n_f)}(\mu)\right)}%
{Z_A^{(n_l)}\left(\alpha_s^{(n_l)}(\mu'),a^{(n_l)}(\mu')\right)}
\zeta_A^0\left(\alpha_{s0}^{(n_f)},a_0^{(n_f)}\right)\,,
\nonumber\\
\zeta_q(\mu',\mu) &=&
\frac{Z_q^{(n_f)}\left(\alpha_s^{(n_f)}(\mu),a^{(n_f)}(\mu)\right)}%
{Z_q^{(n_l)}\left(\alpha_s^{(n_l)}(\mu'),a^{(n_l)}(\mu')\right)}
\zeta_q^0\left(\alpha_{s0}^{(n_f)},a_0^{(n_f)}\right)\,,
\nonumber\\
\zeta_c(\mu',\mu) &=&
\frac{Z_c^{(n_f)}\left(\alpha_s^{(n_f)}(\mu),a^{(n_f)}(\mu)\right)}%
{Z_c^{(n_l)}\left(\alpha_s^{(n_l)}(\mu'),a^{(n_l)}(\mu')\right)}
\zeta_c^0\left(\alpha_{s0}^{(n_f)},a_0^{(n_f)}\right)\,,
\label{Intro:zetaren}
\end{eqnarray}
where $Z_i^{(n_f)}$ are the $\overline{\rm MS}$ renormalization constants in
$n_f$-flavour QCD which we need up to three-loop order.


\section{Calculation}
\label{S:Calc}

Our calculation is automated to a large degree. In a first step we
generate all Feynman diagrams with {\tt QGRAF}~\cite{Nogueira:1991ex}.  
The various diagram topologies are identified and transformed
to {\tt FORM}~\cite{Vermaseren:2000nd} with the help
of {\tt q2e} and {\tt exp}~\cite{Harlander:1997zb,Seidensticker:1999bb}
(these topologies have been investigated in~\cite{Bekavac:2007tk}).
Afterwards we use the
program {\tt FIRE}~\cite{Smirnov:2008iw} to reduce the two-scale three-loop
integrals to four master integrals which can be found in analytic form 
in Ref.~\cite{Bekavac:2009gz}. 

As a cross check we apply the asymptotic expansion (see,
e.g., Ref.~\cite{Smirnov:2002pj}) in the limit $m_c\ll m_b$
and evaluate five expansion terms in $(m_c/m_b)^2$. 
The asymptotic expansion is automated in the program {\tt exp} which provides
output that is passed to the 
package {\tt MATAD}~\cite{Steinhauser:2000ry} performing the actual
calculation.

In the following we present explicit results for the two-point functions
and $\Gamma_{A\bar{c}c}$ needed for the construction of the decoupling constants.
Other vertex functions can be easily reconstructed from the bare decoupling
coefficient $\zeta_{\alpha_s}^0$ in Section~\ref{S:as}
(see Eq.~(\ref{Intro:zetaparams0})).

\subsection{Gluon self-energy}

The bare gluon self-energy at $q^2=0$ in the full theory can be cast in the
following form\footnote{Note that $\Gamma(\varepsilon)=1/\varepsilon + {\cal
    O}(1)$.} 
\begin{eqnarray}
\Pi_A(0) &=&  \frac{1}{3}
\left( n_b m_{b0}^{-2\varepsilon} + n_c m_{c0}^{-2\varepsilon} \right)
T_F \frac{\alpha_{s0}^{(n_f)}}{\pi} \Gamma(\varepsilon)
\nonumber\\
&&{} + P_h \left( n_b m_{b0}^{-4\varepsilon} + n_c m_{c0}^{-4\varepsilon} \right)
T_F \left(\frac{\alpha_{s0}^{(n_f)}}{\pi} \Gamma(\varepsilon)\right)^2
\nonumber\\
&&{} + \biggl[ \left(P_{hg} + P_{hl} T_F n_l\right)
\left( n_b m_{b0}^{-6\varepsilon} + n_c m_{c0}^{-6\varepsilon} \right)
+ P_{hh} T_F
\left( n_b^2 m_{b0}^{-6\varepsilon} + n_c^2 m_{c0}^{-6\varepsilon} \right)
\nonumber\\
&&\hphantom{{}+\biggl[\biggr.}
+ P_{bc}\left(\frac{m_{c0}}{m_{b0}}\right)
T_F n_b n_c \left(m_{b0} m_{c0}\right)^{-3\varepsilon} \biggr]
T_F \left(\frac{\alpha_{s0}^{(n_f)}}{\pi} \Gamma(\varepsilon)\right)^3
+ \cdots
\label{Calc:Pi0}
\end{eqnarray}
where the exact dependence on $\varepsilon=(4-d)/2$ ($d$ is the space-time
dimension) of the bare two-loop result is given by 
\begin{equation}
P_h = \frac{1}{4 (2-\varepsilon) (1+2\varepsilon)} \left[
- C_F \frac{\varepsilon}{3}  (9+7\varepsilon-10\varepsilon^2)
+ C_A \frac{3+11\varepsilon-\varepsilon^2-15\varepsilon^3+4\varepsilon^5}%
{2 (1-\varepsilon) (3+2\varepsilon)} \right]
\label{Calc:Ph}
\end{equation}
($C_F=(N_C^2-1)/(2N_C)$ and $C_A=N_C$ are the eigenvalues of
the quadratic Casimir operators of the fundamental and adjoint 
representation of $SU(N_C)$, respectively, and $T_F=1/2$
is the index of the fundamental representation).
The three-loop quantities $P_{hg}$, $P_{hl}$ and $P_{hh}$ are only available as an
expansion in $\varepsilon$. The analytic results read
\begin{eqnarray}
P_{hg} &=& C_F^2 \frac{\varepsilon^2}{24} \left[ 17
- \frac{1}{8} \left( 95 \zeta_3 + \frac{274}{3} \right) \varepsilon + \cdots \right]
\nonumber\\
&&{} - C_F C_A \frac{\varepsilon}{288} \left[ 89
- \left( 36 \zeta_3 - \frac{785}{6} \right) \varepsilon
- 9 \left( 4 B_4 - \frac{\pi^4}{5} + \frac{1957}{24} \zeta_3 - \frac{10633}{162} \right)
\varepsilon^2 + \cdots \right]
\nonumber\\
&&{} + \frac{C_A^2}{1152} \Biggl[ 3 \xi + 41
- \frac{1}{2} \left( 21 \xi - \frac{781}{3} \right) \varepsilon
- \left( 108 \zeta_3 - \frac{137}{4} \xi - \frac{3181}{12} \right) \varepsilon^2
\nonumber\\
&&\hphantom{{}+\frac{C_A^2}{1152}\Biggl[\Biggr.}
\!- \left( 72 B_4 - \frac{27}{5} \pi^4 - \left( 24 \xi - \frac{1805}{4} \right) \zeta_3
+ \frac{1}{24} \left( 3577 \xi + \frac{42799}{9} \right) \right) \varepsilon^3
+ \cdots \Biggr]\,,
\nonumber\\
P_{hl} &=& \frac{5}{72} C_F \varepsilon \left[ 1 - \frac{31}{30} \varepsilon
+ \frac{971}{180} \varepsilon^2 + \cdots \right]
\nonumber\\
&&{} - \frac{C_A}{72} \left[ 1 + \frac{5}{6} \varepsilon + \frac{101}{12} \varepsilon^2
+ \left( 8 \zeta_3 - \frac{3203}{216} \right) \varepsilon^3 + \cdots \right]\,,
\nonumber\\
P_{hh} &=& C_F \frac{\varepsilon}{18} \left[ 1 - \frac{5}{6} \varepsilon
+ \frac{1}{32} \left( 63 \zeta_3 + \frac{218}{9} \right) \varepsilon^2 + \cdots \right]
\nonumber\\
&&{} - \frac{C_A}{144} \left[ 1 + \frac{35}{6} \varepsilon + \frac{37}{12} \varepsilon^2
- \frac{1}{8} \left( 287 \zeta_3 - \frac{6361}{27} \right) \varepsilon^3 + \cdots \right]\,,
\label{Calc:P3}
\end{eqnarray}
where $\xi=1-a_0^{(n_f)}$,  and~\cite{Broadhurst:1991fi}
\[
B_4 = 16 \Li4\left(\frac{1}{2}\right) + \frac{2}{3} \log^2 2 (\log^2 2 - \pi^2)
- \frac{13}{180} \pi^4\,.
\]

A new result obtained in this paper is the analytic expression for 
$P_{bc}(x)$ which arises from diagrams where $b$ and $c$ quarks are 
simultaneously present 
in the loops (see Fig.~\ref{F:Glue} for typical diagrams). 
The analytic expression is given by
\begin{eqnarray}
P_{bc}(x) &=& C_F \frac{\varepsilon}{9} \left[ 1 - \frac{5}{6} \varepsilon
+ p_F(x) \varepsilon^2 + \cdots \right]
\nonumber\\
&&{} - \frac{C_A}{72} \left[ 1 + \frac{35}{6} \varepsilon
+ \left( \frac{9}{2} L^2 + \frac{37}{12} \right) \varepsilon^2
+  p_A(x) \varepsilon^3 + \cdots \right]\,,
\label{Calc:Pbc}
\end{eqnarray}
with $L=\log x$,
\begin{eqnarray*}
p_F(x) &=& \frac{9}{128} \Biggl[
\frac{(1+x^2)(5-2x^2+5x^4)}{x^3} L_-(x)\\
&&{} - \frac{5-38x^2+5x^4}{x^2} L^2
+ 10 \frac{1-x^4}{x^2} L
- 10 \frac{(1-x^2)^2}{x^2} \Biggr]
+ \frac{109}{144}\,,\\
p_A(x) &=& 24 L_+(x)
- \frac{3}{4} \frac{(1+x^2)(4+11x^2+4x^4)}{x^3} L_-(x)\\
&&{} + \frac{(1+6x^2)(6+x^2)}{2x^2} L^2
- 6 \frac{1-x^4}{x^2} L
+ 6 \frac{(1-x^2)^2}{x^2}
+ 8 \zeta_3 + \frac{6361}{216}\,,
\end{eqnarray*}
where the functions $L_\pm(x)$ are defined in~(\ref{Ix:L}).
The function $P_{bc}(x)$ satisfies the properties
\begin{equation}
P_{bc}(x^{-1}) = P_{bc}(x)\,,\quad
P_{bc}(1) = 2 P_{hh}\,,
\label{Calc:testPi}
\end{equation}
which are a check of our result.
For $x\to0$, the hard contribution to $P_{bc}(x) x^{-3\varepsilon}$
is given by $P_{hl}$.
However, there is also a soft contribution, and it is not possible to obtain
a relation between $P_{bc}(x\to0)$ and $P_{hl}$ if they are expanded in $\varepsilon$
(this would be possible for a non-zero $\varepsilon<0$, cf.~(\ref{Ix:0})).

\EPSFIGURE[t]{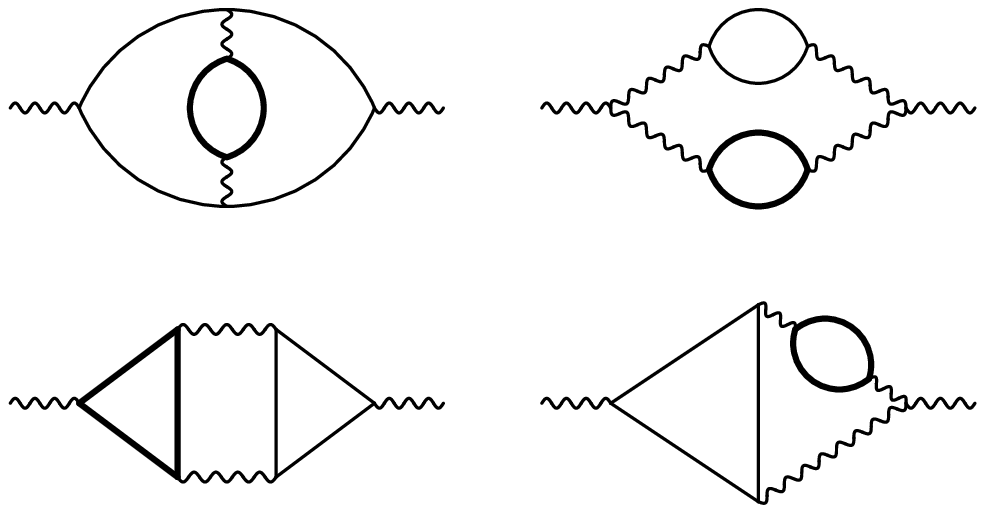}{Feynman diagrams contributing to the gluon self-energy.
Thick and thin straight lines correspond to $b$ and $c$ quarks, respectively.
Wavy lines represent gluons.
\label{F:Glue}}

\FIGURE[t]{\begin{picture}(100,14)
\put(50,7){\makebox(0,0){\includegraphics{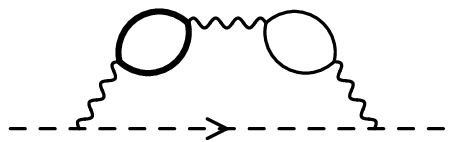}}}
\end{picture}
\caption{Feynman diagrams with two heavy-quark loops 
contributing to the ghost self-energy. The notation is adopted from Fig.~\ref{F:Glue}.
\label{F:Ghost}}}

\subsection{Ghost self-energy}

The bare ghost self-energy at $q^2=0$ can be cast in the form
\begin{eqnarray}
\Pi_c(0) &=&
C_h \left(n_b m_{b0}^{-4\varepsilon} + n_c m_{c0}^{-4\varepsilon}\right)
C_A T_F \left(\frac{\alpha_{s0}^{(n_f)}}{\pi} \Gamma(\varepsilon)\right)^2
\nonumber\\
&&{} + \biggl[ \left(C_{hg} + C_{hl}T_F n_l\right)
\left( n_b m_{b0}^{-6\varepsilon} + n_c m_{c0}^{-6\varepsilon} \right)
+ C_{hh}T_F
\left( n_b^2 m_{b0}^{-6\varepsilon} + n_c^2 m_{c0}^{-6\varepsilon} \right)
\nonumber\\
&&\hphantom{{}+\biggl[\biggr.}
+ C_{bc}\left(\frac{m_{c0}}{m_{b0}}\right)
T_F n_b n_c \left(m_{b0} m_{c0}\right)^{-3\varepsilon} \biggr]
C_A T_F \left(\frac{\alpha_{s0}^{(n_f)}}{\pi} \Gamma(\varepsilon)\right)^3
+ \cdots\,,
\label{Calc:Ghost}
\end{eqnarray}
where the two-loop term is given by
\begin{equation}
C_h = - \frac{(1+\varepsilon) (3-2\varepsilon)}%
{16 (1-\varepsilon) (2-\varepsilon) (1+2\varepsilon) (3+2\varepsilon)}\,,
\label{Calc:Ch}
\end{equation}
and the $\varepsilon$ expansions of the single-scale three-loop coefficients read
\begin{eqnarray}
C_{hg} &=& C_F \frac{\varepsilon}{64} \left[ 5
- \left( 4 \zeta_3 + \frac{9}{2} \right) \varepsilon
- \left( 4 B_4 - \frac{\pi^4}{5} + \frac{57}{2} \zeta_3 - \frac{157}{4} \right)
\varepsilon^2 + \cdots \right]
\nonumber\\
&&{} + \frac{C_A}{2304} \Biggl[ 3 \xi - 47
- \frac{1}{2} \left( 9 \xi + \frac{83}{3} \right) \varepsilon
+ \left( 108 \zeta_3 + \frac{131}{4} \xi - \frac{9083}{36} \right) \varepsilon^2
\nonumber\\
&&\hphantom{{}+\frac{C_A}{2304}\Biggl[\Biggr.}
+ \left( 72 B_4 - \frac{27}{5} \pi^4 + (24 \xi + 407) \zeta_3
- \frac{1}{24} \left( 2239 \xi - \frac{49795}{9} \right) \right)
\varepsilon^3 + \cdots \Biggr]\,,
\nonumber\\
C_{hl} &=& \frac{1}{144} \left[ 1 - \frac{5}{6} \varepsilon + \frac{337}{36} \varepsilon^2
+ \left( 8 \zeta_3 - \frac{5261}{216} \right) \varepsilon^3 + \cdots \right]\,,
\nonumber\\
C_{hh} &=& \frac{1}{72} \left[ 1 - \frac{5}{6} \varepsilon + \frac{151}{36} \varepsilon^2
- \left( 7 \zeta_3 + \frac{461}{216} \right) \varepsilon^3 + \cdots \right]\,.
\label{Calc:C3}
\end{eqnarray}

The function $C_{bc}(x)$ is obtained from the diagram of Fig.~\ref{F:Ghost} and can be
written as
\begin{equation}
C_{bc}(x) = - \frac{3-2\varepsilon}{64(2-\varepsilon)} I(x)\,,
\label{Calc:Cbc}
\end{equation}
with
\begin{equation}
\int \frac{\Pi_b(k^2) \Pi_c(k^2)}{(k^2)^2} d^d k =
i T_F^2 \frac{\alpha_{s0}^2}{16 \pi^\varepsilon} \Gamma^3(\varepsilon)
(m_{b0} m_{c0})^{-3\varepsilon} I\left(\frac{m_{c0}}{m_{b0}}\right)\,,
\label{Calc:Idef}
\end{equation}
where $\Pi_{b}(k^2)$ and $\Pi_{c}(k^2)$ are the $b$- and $c$-loop
contributions to the gluon self-energy. 
The integral $I(x)$ is discussed in Appendix~\ref{S:Ix} where an analytic
result is presented.
In analogy to Eq.~(\ref{Calc:testPi}), we have
\begin{equation}
C_{bc}(x^{-1}) = C_{bc}(x)\,,\quad
C_{bc}(1) = 2 C_{hh}\,.
\end{equation}
For a non-zero $\varepsilon<0$, $C_{bc}(x\to0)\to C_{hl} x^{3\varepsilon}$
(only the hard part survives in~(\ref{Ix:0})).

\subsection{Light-quark self-energy}

The parts of the light-quark self-energy $\Sigma_V(0)$ and $\Sigma_S(0)$
(with vanishing light-quark masses) are conveniently written in the form
\begin{eqnarray}
\Sigma_V(0) &=&
V_h \left(n_b m_{b0}^{-4\varepsilon} + n_c m_{c0}^{-4\varepsilon}\right)
C_F T_F \left(\frac{\alpha_{s0}^{(n_f)}}{\pi} \Gamma(\varepsilon)\right)^2
\nonumber\\
&&{} + \biggl[ \left(V_{hg} + V_{hl} T_F n_l\right)
\left( n_b m_{b0}^{-6\varepsilon} + n_c m_{c0}^{-6\varepsilon} \right)
+ V_{hh} T_F
\left( n_b^2 m_{b0}^{-6\varepsilon} + n_c^2 m_{c0}^{-6\varepsilon} \right)
\nonumber\\
&&\hphantom{{}+\biggl[\biggr.}
+ V_{bc}\left(\frac{m_{c0}}{m_{b0}}\right)
T_F n_b n_c \left(m_{b0} m_{c0}\right)^{-3\varepsilon} \biggr]
C_F T_F \left(\frac{\alpha_{s0}^{(n_f)}}{\pi} \Gamma(\varepsilon)\right)^3
+ \cdots\,,
\nonumber\\
\Sigma_S(0) &=&
S_h \left(n_b m_{b0}^{-4\varepsilon} + n_c m_{c0}^{-4\varepsilon}\right)
C_F T_F \left(\frac{\alpha_{s0}^{(n_f)}}{\pi} \Gamma(\varepsilon)\right)^2
\nonumber\\
&&{} + \biggl[ \left(S_{hg} + S_{hl} T_F n_l\right)
\left( n_b m_{b0}^{-6\varepsilon} + n_c m_{c0}^{-6\varepsilon} \right)
+ S_{hh} T_F
\left( n_b^2 m_{b0}^{-6\varepsilon} + n_c^2 m_{c0}^{-6\varepsilon} \right)
\nonumber\\
&&\hphantom{{}+\biggl[\biggr.}
+ S_{bc}\left(\frac{m_{c0}}{m_{b0}}\right)
T_F n_b n_c \left(m_{b0} m_{c0}\right)^{-3\varepsilon} \biggr]
C_F T_F \left(\frac{\alpha_{s0}^{(n_f)}}{\pi} \Gamma(\varepsilon)\right)^3
+ \cdots\,,
\label{Calc:Sigma}
\end{eqnarray}
where
\begin{equation}
V_h =  - \frac{\varepsilon (1+\varepsilon) (3-2\varepsilon)}%
{8 (1-\varepsilon) (2-\varepsilon) (1+2\varepsilon) (3+2\varepsilon)}\,,\quad
S_h = - \frac{(1+\varepsilon) (3-2\varepsilon)}%
{8 (1-\varepsilon) (1+2\varepsilon) (3+2\varepsilon)}\,,
\label{Calc:VSh}
\end{equation}
and
\begin{eqnarray}
V_{hg} &=& - C_F \frac{\varepsilon}{96} \left[ 1 - \frac{39}{2} \varepsilon
+ \left( 12 \zeta_3 + \frac{335}{12} \right) \varepsilon^2 + \cdots \right]
\nonumber\\
&&{} + \frac{C_A}{192} \Biggl[ \xi - 1
- \left( 3 \xi + \frac{10}{3} \right) \varepsilon
+ \frac{1}{3} \left( 35 \xi - \frac{227}{3} \right) \varepsilon^2
\nonumber\\
&&\hphantom{{}+\frac{C_A}{192}\Biggl[\Biggr.}
+ \left( 8 (\xi + 2) \zeta_3
- \frac{1}{9} \left( 407 \xi - \frac{1879}{6} \right)
\right) \varepsilon^3 + \cdots \Biggr]\,,
\nonumber\\
V_{hl} &=& \frac{\varepsilon}{72} \left[ 1 - \frac{5}{6} \varepsilon
+ \frac{337}{36} \varepsilon^2 + \cdots \right]\,,
\nonumber\\
V_{hh} &=& \frac{\varepsilon}{36} \left[ 1 - \frac{5}{6} \varepsilon
+ \frac{151}{36} \varepsilon^2 + \cdots \right]\,,
\nonumber\\
S_{hg} &=& C_F \frac{\varepsilon}{16} \left[ 5
- \left( 4 \zeta_3 + \frac{23}{3} \right) \varepsilon
-  \left( 4 B_4 - \frac{\pi^4}{5} + \frac{53}{2} \zeta_3 - \frac{257}{6} \right)
\varepsilon^2 + \cdots \right]
\nonumber\\
&&{} + \frac{C_A}{576} \Biggl[ - 3 \xi - 41
+ \left( 9 \xi - \frac{124}{3} \right) \varepsilon
+ \left( 144 \zeta_3 - 35 \xi - \frac{836}{9} \right) \varepsilon^2
\nonumber\\
&&\hphantom{{}+\frac{C_A}{576}\Biggl[\Biggl.}
+ \left( 72 B_4 - \frac{36}{5} \pi^4 - (24 \xi - 581) \zeta_3
+ \frac{1}{3} \left( 407 \xi - \frac{9751}{9} \right) \right) \varepsilon^3
+ \cdots \Biggr]\,,
\nonumber\\
S_{hl} &=& \frac{1}{36} \left[ 1 - \frac{4}{3} \varepsilon
+ \frac{88}{9} \varepsilon^2
+ 8 \left( \zeta_3 - \frac{98}{27} \right) \varepsilon^3
+ \cdots \right]\,,
\nonumber\\
S_{hh} &=& \frac{1}{18} \left[ 1 - \frac{4}{3} \varepsilon
+ \frac{83}{18} \varepsilon^2
- \left( 7 \zeta_3 + \frac{457}{108} \right) \varepsilon^3
+ \cdots \right]\,.
\label{Calc:V3}
\end{eqnarray}
Exact $d$-dimensional expressions for these coefficients have been obtained
in~\cite{Grozin:2006xm}.

The quantities
$V_{bc}(x)$ and $S_{bc}(x)$ arise from diagrams similar to Fig.~\ref{F:Ghost}
and can be expressed in terms of $I(x)$:
\begin{equation}
V_{bc}(x) = - \frac{\varepsilon(3-2\varepsilon)}{32(2-\varepsilon)} I(x)\,,\quad
S_{bc}(x) = - \frac{3-2\varepsilon}{32} I(x)\,.
\label{Calc:VSbc}
\end{equation}
They satisfy the relations analogous to Eq.~(\ref{Calc:testPi}) which again serves
as a welcome check of our calculation.
Retaining only the hard part of~(\ref{Ix:0}) for $x\to0$,
we reproduce $V_{hl}$, $S_{hl}$.
$V_{bc}$ has been calculated up to $\mathcal{O}(\varepsilon^3)$ in
Ref.~\cite{Bekavac:2009zc}.

\subsection{Ghost--gluon vertex}

The two-loop correction vanishes in the arbitrary covariant gauge
exactly in $\varepsilon$, see Appendix~\ref{S:Ghost}.
For the same reasons, the three-loop correction contains only diagrams
with a single quark loop (bottom or charm),
and vanishes in Landau gauge:
\begin{eqnarray}
&&\Gamma_{A\bar{c}c} = 1 + \Gamma_3 (1-\xi)
(n_b m_{b0}^{-6\varepsilon} + n_c m_{c0}^{-6\varepsilon}) C_A^2 T_F
\left(\frac{\alpha_{s0}^{(n_f)}}{\pi} \Gamma(\varepsilon)\right)^3 + \cdots\,,
\label{Calc:GAcc}\\
&&\Gamma_3 = - \frac{1}{384} \left[ 1
- \frac{5}{2} \varepsilon
+ \frac{67}{6} \varepsilon^2
+ \left(8 \zeta_3 - \frac{727}{18}\right) \varepsilon^3
+ \cdots \right]\,.
\nonumber
\end{eqnarray}


\section{Decoupling for $\alpha_s$}
\label{S:as}

The gauge parameter
dependence cancels in the bare decoupling constant~(\ref{Intro:zetaparams0})
(which relates $\alpha_{s0}^{(n_l)}$ to $\alpha_{s0}^{(n_f)}$, 
see Eq.~(\ref{Intro:params0})). Since the result is more compact we
present analytical expressions for $\left(\zeta_{\alpha_s}^0\right)^{-1}$
which reads
\begin{eqnarray}
\left(\zeta_{\alpha_s}^0\right)^{-1} &=& 1
+ \frac{1}{3} \left(n_b m_{b0}^{-2\varepsilon} + n_c m_{c0}^{-2\varepsilon}\right)
T_F \frac{\alpha_{s0}^{(n_f)}}{\pi} \Gamma(\varepsilon)
\nonumber\\
&&{} + Z_h \varepsilon T_F
(n_b m_{b0}^{-4\varepsilon} + n_c m_{c0}^{-4\varepsilon})
\left(\frac{\alpha_{s0}^{(n_f)}}{\pi} \Gamma(\varepsilon)\right)^2
\nonumber\\
&&{} + \biggl[ \left(Z_{hg} + Z_{hl} T_F n_l\right)
\left( n_b m_{b0}^{-6\varepsilon} + n_c m_{c0}^{-6\varepsilon} \right)
+ Z_{hh} T_F
\left( n_b^2 m_{b0}^{-6\varepsilon} + n_c^2 m_{c0}^{-6\varepsilon} \right)
\nonumber\\
&&\hphantom{{}+\biggl[\biggr.}
+ Z_{bc}\left(\frac{m_{c0}}{m_{b0}}\right)
T_F n_b n_c \left(m_{b0} m_{c0}\right)^{-3\varepsilon} \biggr] \varepsilon
T_F \left(\frac{\alpha_{s0}^{(n_f)}}{\pi} \Gamma(\varepsilon)\right)^3
+ \cdots\,,
\label{as:bare}
\end{eqnarray}
where
\begin{eqnarray*}
Z_h &=& \frac{1}{4 (2-\varepsilon) (1+2\varepsilon)} \left[
- \frac{1}{3} C_F (9+7\varepsilon-10\varepsilon^2)
+ \frac{1}{2} C_A
\frac{10+11\varepsilon-4\varepsilon^2-4\varepsilon^3}{3+2\varepsilon}
\right]\,,\\
Z_{hg} &=& \frac{C_F^2 \varepsilon}{24}
\left[17 - \frac{1}{4} \left(\frac{95}{2} \zeta_3 + \frac{137}{3}\right) \varepsilon
+ \cdots\right]\\
&&{} - \frac{C_F C_A}{72}
\left[11 + \frac{257}{6} \varepsilon
- \frac{1}{16} \left(\frac{3819}{2} \zeta_3 - \frac{8549}{9}\right) \varepsilon^2
+ \cdots\right]\\
&&{} + \frac{C_A^2}{216}
\left[19 + \frac{359}{24} \varepsilon
+ \frac{1}{32} \left(\frac{45}{2} \zeta_3 - \frac{3779}{3}\right) \varepsilon^2
+ \cdots\right]\,,\\
Z_{hl} &=& \frac{C_F}{72}
\left[5 - \frac{31}{6} \varepsilon + \frac{971}{36} \varepsilon^2 + \cdots\right]
- \frac{C_A}{216}
\left[5 - \frac{17}{6} \varepsilon + \frac{343}{12} \varepsilon^2 + \cdots\right]\,,\\
Z_{hh} &=& \frac{C_F}{18}
\left[1 - \frac{5}{6} \varepsilon
+ \frac{1}{16} \left(\frac{63}{2} \zeta_3 + \frac{109}{9}\right) \varepsilon^2
+ \cdots\right]\\
&&{} - \frac{C_A}{108}
\left[5 - \frac{113}{24} \varepsilon
- \frac{1}{16} \left(\frac{189}{2} \zeta_3 - 311\right) \varepsilon^2
+ \cdots\right]\,,\\
Z_{bc}(x) &=& \frac{C_F}{9} \left[1 - \frac{5}{6} \varepsilon
+ z_F(x) \varepsilon^2 + \cdots\right]
- \frac{C_A}{54} \left[5 - \frac{113}{24} \varepsilon
+ z_A(x) \varepsilon^2 + \cdots\right]\,,\\
z_F(x) &=& \frac{9}{64} \Biggl[
\frac{(1+x^2) (5-2x^2+5x^4)}{2 x^3} L_-(x)\\
&&{} - \frac{5-38x^2+5x^4}{2 x^2} L^2
+ 5 \frac{1-x^4}{x^2} L
- 5 \frac{(1-x^2)^2}{x^2} \Biggr]
+ \frac{109}{144}\,,\\
z_A(x) &=& \frac{3}{16} \Biggl[
- 9 \frac{(1+x^2) (1+x^4)}{2 x^3} L_-(x)\\
&&{} + \frac{9+92x^2+9x^4}{2 x^2} L^2
- 9 \frac{1-x^4}{x^2} L
+ 9 \frac{(1-x^2)^2}{x^2} \Biggr]
+ \frac{311}{16}\,.
\end{eqnarray*}
Note that $Z_{bc}(x^{-1})=Z_{bc}(x)$, $Z_{bc}(1)=2Z_{hh}$.
If desired, the vertices $\Gamma_{A\bar{q}q}$ and $\Gamma_{AAA}$ can be reconstructed
using Eq.~(\ref{Intro:zetaparams0}).

In order to relate the renormalized couplings $\alpha_s^{(n_f)}(\mu)$ and  $\alpha_s^{(n_l)}(\mu)$,
we first express all bare quantities in the right-hand side of the equation
\[
\alpha_{s0}^{(n_l)} = \zeta_{\alpha_s}^0(\alpha_{s0}^{(n_f)},m_{b0},m_{c0}) \alpha_{s0}^{(n_f)}
\]
via the $\overline{\mbox{MS}}$ renormalized
ones~\cite{vanRitbergen:1997va,Czakon:2004bu,Chetyrkin:1997dh,Vermaseren:1997fq}
\begin{eqnarray}
&&\frac{\alpha_{s0}^{(n_f)}}{\pi} \Gamma(\varepsilon) =
\frac{\alpha_s^{(n_f)}(\mu)}{\pi \varepsilon}
Z_\alpha^{(n_f)}\left(\alpha_s^{(n_f)}(\mu)\right)
e^{\gamma_E \varepsilon} \Gamma(1+\varepsilon)
\mu^{2\varepsilon}\,,
\label{as:MSbar}\\
&&m_{b0} = Z_m^{(n_f)}\left(\alpha_s^{(n_f)}(\mu)\right) m_b(\mu)
\label{as:massren}
\end{eqnarray}
(and similarly for $m_{c0}$).
This leads to an equation where 
$\alpha_{s0}^{(n_l)}$ is expressed via the $n_f$-flavour
$\overline{\mbox{MS}}$ renormalized quantities%
\footnote{Note that the masses $m_c(\mu)$ and $m_b(\mu)$ (and $m_{c0}$, $m_{b0}$)
are those in the full $n_f$-flavour QCD.
They do not exist in the low-energy $n_l$-flavour QCD,
and therefore we do not assign a superscript $n_f$ to these masses.}
$\alpha_s^{(n_f)}(\mu)$, $m_c(\mu)$ and $m_b(\mu)$.
In a next step we invert the series
\[
\frac{\alpha_{s0}^{(n_l)}}{\pi} \Gamma(\varepsilon) =
\frac{\alpha_s^{(n_l)}(\mu')}{\pi \varepsilon}
Z_\alpha^{(n_l)}\left(\alpha_s^{(n_l)}(\mu')\right)
e^{\gamma_E \varepsilon} \Gamma(1+\varepsilon)
\left(\mu^\prime\right)^{2\varepsilon}
\]
to express $\alpha_s^{(n_l)}(\mu')$ via $\alpha_{s0}^{(n_l)}$,
and substitute the series for $\alpha_{s0}^{(n_l)}$ derived above.

In order to obtain compact formulae it is convenient to set
$\mu=\bar{m}_b$ where $\bar{m}_b$ is defined as the root of the equation
$m_b(\bar{m}_b) = \bar{m}_b$.
Furthermore, we choose $\mu'=m_c(\bar{m}_b)$ and thus obtain
$\alpha_s^{(n_l)}(m_c(\bar{m}_b))$ as a series in $\alpha_s^{(n_f)}(\bar{m}_b)$
with coefficients depending on
\begin{equation}
x = \frac{m_c(\bar{m}_b)}{\bar{m}_b}\,.
\label{as:x}
\end{equation}
We obtain  ($L=\log x$)
\begin{equation}
\zeta_{\alpha_s}(m_c(\bar{m}_b),\bar{m}_b)\!=\!
e^{-2L\varepsilon} \left[ 1
+ d_1 \frac{\alpha_s^{(n_f)}(\bar{m}_b)}{\pi}
+ d_2 \left(\frac{\alpha_s^{(n_f)}(\bar{m}_b)}{\pi}\right)^2
+ d_3 \left(\frac{\alpha_s^{(n_f)}(\bar{m}_b)}{\pi}\right)^3\!\! + \cdots
\right],
\label{as:renorm}
\end{equation}
where
\begin{eqnarray*}
d_1 &=& - \left[ 11 C_A - 4 T_F (n_l + n_c) \right] \frac{L}{6}
+ \left\{  \left[ 11 C_A - 4 T_F (n_l + n_c) \right] L^2
  - T_F (n_b + n_c) \frac{\pi^2}{6} \right\} \frac{\varepsilon}{6}\\
&&{} - \left\{ \left[ 11 C_A - 4 T_F (n_l + n_c) \right] L^3
  - T_F n_c \frac{\pi^2}{2} L - T_F (n_b + n_c) \zeta_3 \right\} \frac{\varepsilon^2}{9}
+ \mathcal{O}(\varepsilon^3)\,,\\
d_2 &=& \left[11 C_A - 4 T_F (n_l+n_c) \right]^2 \frac{L^2}{36}
- \left[ 17 C_A^2 - 6 C_F T_F (n_l - n_c) - 10 C_A T_F (n_l + n_c) \right] \frac{L}{12}\\
&&{} - \frac{(39 C_F - 32 C_A) T_F (n_b+n_c)}{144}\\
&&{} + \biggl\{
- \left[11 C_A - 4 T_F (n_l+n_c) \right]^2 \frac{L^3}{18}\\
&&\hphantom{{}+\biggl\{\biggr.}
+ \left[ 17 C_A^2 - 6 C_F T_F (n_l - 2 n_c) - 10 C_A T_F (n_l + n_c) \right] \frac{L^2}{6}\\
&&\hphantom{{}+\biggl\{\biggr.}
+ T_F \left[ \frac{13}{12} C_F n_c
  + \frac{C_A}{9} \left( \frac{11}{12} \pi^2 (n_b + n_c) - 8 n_c \right) 
- T_F \frac{\pi^2}{27} (n_b + n_c)(n_l + n_c)
\right] L\\
&&\hphantom{{}+\biggl\{\biggr.}
+ \left[ \frac{C_F}{4} \left( \pi^2 + \frac{35}{2} \right)
  - \frac{C_A}{3} \left( \frac{5}{4} \pi^2 + \frac{43}{3} \right) \right]
\frac{T_F (n_b + n_c)}{12}
\biggr\} \varepsilon + \mathcal{O}(\varepsilon^2)\,,\\
d_3 &=& - \frac{\left[11 C_A - 4 T_F (n_l+n_c) \right]^3}{216} L^3\\
&&{} + \biggl[ \frac{935}{24} C_A^3
- \frac{55}{4} C_F C_A T_F (n_l-n_c)
- \frac{445}{12} C_A^2 T_F (n_l+n_c)\\
&&\hphantom{{}+\biggl[\biggr.}
+ 5 C_F T_F^2 (n_l^2 - n_c^2)
+ \frac{25}{3} C_A T_F^2 (n_l+n_c)^2
\biggr] \frac{L^2}{6}\\
&&{} + \biggl[ - \frac{2857}{1728} C_A^3
- C_F^2 T_F \frac{n_l - 9 n_c}{16}
+ \frac{C_F C_A T_F}{48} \left( \frac{205}{6} n_l - 19 n_c + \frac{143}{3} n_b \right)\\
&&\hphantom{{}+\biggl[\biggr.}
+ \frac{C_A^2 T_F}{27} \left( \frac{1415}{32} n_l + \frac{359}{32} n_c - 22 n_b \right)
- C_F T_F^2 \frac{(n_l + n_c) (11 n_l + 30 n_c) + 26 n_l n_b}{72}\\
&&\hphantom{{}+\biggl[\biggr.}
- C_A T_F^2 \frac{(n_l + n_c) (79 n_l - 113 n_c) - 128 n_l n_b}{432}
\biggr] L\\
&&{} + \biggl[
\frac{C_F^2}{96} \left( \frac{95}{2} \zeta_3 - \frac{97}{3} \right)
- \frac{C_F C_A}{96} \left( \frac{1273}{8} \zeta_3 - \frac{2999}{27} \right)
- \frac{C_A^2}{768} \left( \frac{5}{2} \zeta_3 - \frac{11347}{27} \right)\\
&&\hphantom{{}+\biggl[\biggr.}
- \frac{41}{162} C_F T_F n_l
- \frac{C_F T_F (n_b+n_c)}{16} \left( \frac{7}{4} \zeta_3 - \frac{103}{81} \right)\\
&&\hphantom{{}+\biggl[\biggr.}
- \frac{C_A T_F n_l}{2592}
- \frac{7}{64} C_A T_F (n_b+n_c) \left( \frac{1}{2} \zeta_3 - \frac{35}{81} \right)
\biggr] T_F (n_b+n_c)\\
&&{} + T_F^2 n_b n_c \left( C_F d_F(x) + C_A d_A(x) \right)
+ \mathcal{O}(\varepsilon)\,.
\end{eqnarray*}
The functions
\begin{eqnarray*}
d_F(x) &=& - \frac{(1 + x^2) (5 - 2 x^2 + 5 x^4)}{128 x^3} L_-(x)
+ \frac{7}{32} \zeta_3\\
&&{} + \left[ \frac{5}{4} \frac{(1-x^2)^2}{x^2} + \frac{11}{3} \right] \frac{L^2}{32}
- \frac{5}{4} \left[ \frac{1-x^4}{16 x^2} + \frac{1}{3} \right] L
+ \frac{5}{64} \frac{(1-x^2)^2}{x^2}\,,\\
d_A(x) &=& - \frac{(1+x^2) (1+x^4)}{64 x^3} L_-(x)
+ \frac{7}{64} \zeta_3\\
&&{} + \left[ \frac{(1-x^2)^2}{2 x^2} + \frac{5}{3} \right] \frac{L^2}{32}
- \left[ \frac{1-x^4}{2 x^2} - \frac{113}{27} \right] \frac{L}{16}
+ \frac{(1-x^2)^2}{32 x^2}
\end{eqnarray*}
are defined in such a way that $d_{F,A}(1)=0$.
Thus for $x=1$ Eq.~(\ref{as:renorm})
reduces to the ordinary decoupling of $n_b+n_c$ flavours
with the same mass~\cite{Chetyrkin:1997un}.
For $x\ll1$ the functions $d_{F}(x)$ and $d_{A}(x)$ become
\begin{eqnarray}
d_F(x) &=& - \frac{1}{36} \left( 13 L - \frac{89}{12} \right) + \frac{7}{32} \zeta_3
+ \left( 2 L + \frac{13}{30} \right) \frac{x^2}{15} + \cdots
\nonumber\\
d_A(x) &=& \frac{1}{27} \left( 8 L - \frac{41}{16} \right) + \frac{7}{64} \zeta_3
- \left( \frac{1}{2} L^2 - \frac{121}{30} L + \frac{19}{225} \right)
\frac{x^2}{60} + \cdots\,.
\label{as:x0}
\end{eqnarray}

An expression for $\alpha_s^{(n_f)}(\bar{m}_b)$ via $\alpha_s^{(n_l)}(m_c(\bar{m}_b))$
can be obtained by inverting the series~(\ref{as:renorm}).
If one wants to express $\alpha_s^{(n_l)}(\mu_c)$ as a truncated series
in $\alpha_s^{(n_f)}(\mu_b)$ (without resummation) for some other choice
of $\mu_b\sim m_b$ and $\mu_c\sim m_c$, this can be easily done in three steps:
$(i)$ run from $\mu_b$ to $\bar{m}_b$ in the $n_f$-flavour theory (without resummation);
$(ii)$ use Eq.~(\ref{as:renorm}) for the decoupling;
and $(iii)$ run from $m_c(\bar{m}_b)$ to $\mu_c$ in the $n_l$-flavour theory
(without resummation).
After that, relating $\alpha_s^{(n_l)}(\mu')$ and $\alpha_s^{(n_f)}(\mu)$
for any values of $\mu$ and $\mu'$ (possibly widely separated from $m_b$ and $m_c$)
can be done in a similar way:
$(i)$ run from $\mu$ to $\mu_b$ in the $n_f$-flavour theory (with resummation);
$(ii)$ use the decoupling relation derived above;
and $(iii)$ run from $\mu_c$ to $\mu'$ in the $n_l$-flavour theory
(with resummation).
The steps $(i)$ and $(iii)$ can conveniently be performed using the
program {\tt RunDec}~\cite{Chetyrkin:2000yt}.

In the case of QCD ($T_F=1/2$, $C_A=3$, $C_F=4/3$, $n_b=n_c=1$)
the decoupling constant in Eq.~(\ref{as:renorm}) reduces to (for
  $\varepsilon=0$)
\begin{eqnarray}
&&\zeta_{\alpha_s}(m_c(\bar{m}_b),\bar{m}_b) = 1
+ \frac{2 n_l - 31}{6} L \frac{\alpha_s^{(n_f)}(\bar{m}_b)}{\pi}
\nonumber\\
&&{} + \left[
\frac{(2 n_l - 31)^2}{36} L^2
+ \frac{19 n_l - 142}{12} L
+ \frac{11}{36} \right]
\left(\frac{\alpha_s^{(n_f)}(\bar{m}_b)}{\pi}\right)^2
\nonumber\\
&&{} + \biggl[
\frac{(2 n_l - 31)^3}{216} L^3
+ \left( \frac{95}{9} n_l^2 - \frac{485}{2} n_l
+ \frac{58723}{48} \right) \frac{L^2}{8}
\nonumber\\
&&\hphantom{{}+\biggl\{\biggr.}
- \left( \frac{325}{6} n_l^2 - \frac{15049}{6} n_l + 12853 \right) \frac{L}{288}
- \frac{(1+x^2) (19 - 4 x^2 + 19 x^4)}{768 x^3} L_-(x)
\nonumber\\
&&\hphantom{{}+\biggl\{\biggr.}
+ \frac{19}{768} \left( \frac{(1 - x^2)^2}{x^2} (L^2 + 2) - 2 \frac{1 - x^4}{x^2} L \right)
\nonumber\\
&&\hphantom{{}+\biggl\{\biggr.}
- \frac{1}{1728} \left( \frac{82043}{8} \zeta_3 + \frac{2633}{9} n_l
- \frac{572437}{36} \right) \biggr]
\left(\frac{\alpha_s^{(n_f)}(\bar{m}_b)}{\pi}\right)^3 + \cdots\,.
\label{as:su3}
\end{eqnarray}
For $x\ll1$ the coefficient of $(\alpha_s/\pi)^3$ becomes
\begin{eqnarray*}
&&\frac{(2 n_l - 31)^3}{216} L^3
+ \frac{5 (2 n_l - 31) (19 n_l - 142)}{144} L^2
- \frac{325 n_l^2 - 15049 n_l + 77041}{1728} L\\
&&{} - \frac{1}{1728} \left( \frac{82043}{8} \zeta_3 + \frac{2633}{9} n_l
- \frac{563737}{36} \right)
- \left( L^2 - \frac{683}{45} L - \frac{926}{675} \right) \frac{x^2}{160}
+ \mathcal{O}(x^4)\,.
\end{eqnarray*}


\section{Decoupling for the light-quark masses}
\label{S:m}

The bare quark mass decoupling coefficient $\zeta_m^0$ of Eq.~(\ref{Intro:zetafields0})
is determined by $\Sigma_{V}(0)$ and $\Sigma_{S}(0)$, see Eq.~(\ref{Calc:Sigma});
it is gauge parameter independent.
The renormalized decoupling constant $\zeta_m$ in Eq.~(\ref{Intro:zetaren})
(see~\cite{Chetyrkin:1997dh,Vermaseren:1997fq} for the mass renormalization constants)
can be obtained by re-expressing $\alpha_s^{(n_l)}$ in the denominator via
$\alpha_s^{(n_f)}$ 
(cf. Sect.~\ref{S:as}; note that in $\zeta_{\alpha_s}$
positive powers of $\varepsilon$ should be kept). 
Our result reads
\begin{equation}
\zeta_m(m_c(\bar{m}_b),\bar{m}_b) = 1
+ d^m_1 C_F \frac{\alpha_s^{(n_f)}(\bar{m}_b)}{\pi}
+ d^m_2 C_F \left(\frac{\alpha_s^{(n_f)}(\bar{m}_b)}{\pi}\right)^2
+ d^m_3 C_F \left(\frac{\alpha_s^{(n_f)}(\bar{m}_b)}{\pi}\right)^3
+ \cdots
\,,
\label{m:ren}
\end{equation}
where
\begin{eqnarray*}
d^m_1 &=& - \frac{3}{2} L
\left( 1 - L \varepsilon + \frac{2}{3} L^2 \varepsilon^2
+ \mathcal{O}(\varepsilon^3) \right)\,,\\
d^m_2 &=&
\left[ 9 C_F + 11 C_A - 4 T_F (n_l + n_c) \right] \frac{L^2}{8}
- \left[ 9 C_F + 97 C_A - 20 T_F (n_l + n_c) \right] \frac{L}{48}\\
&&{} + \frac{89}{288} T_F (n_b + n_c)\\
&&{} + \biggl\{
- \left[9 C_F + 11 C_A - 4 T_F (n_l + n_c)\right] \frac{L^3}{4}
+\left[9 C_F + 97 C_A - 20 T_F (n_l + n_c)\right] \frac{L^2}{24}
\nonumber\\
&&\hphantom{{}+\biggl\{\biggr.}
+ \frac{3 \pi^2 n_b - 89 n_c}{72} T_F L
- \left( 5 \pi^2 + \frac{869}{6} \right) T_F\frac{n_b + n_c}{288}
\biggr\} \varepsilon 
+ \mathcal{O}(\varepsilon^2)
\,,\\
d^m_3 &=&
\biggl[
- \frac{(9 C_F + 11 C_A) (9 C_F + 22 C_A)}{16}
+ \frac{27 C_F + 44 C_A}{4} T_F (n_l + n_c)\\
&&\hphantom{\biggl[\biggr.}
- T_F^2 \bigl( 2 (n_l + n_c)^2 - n_b n_c \bigr)
\biggr] \frac{L^3}{9}\\
&&{} + \biggl[
\frac{9}{4} C_F^2 + 27 C_F C_A + \frac{1373}{36} C_A^2
- \left( 9 C_F + \frac{197}{9} C_A \right) T_F (n_l + n_c)\\
&&\hphantom{{}+\biggl[\biggr.}
+ T_F^2 \frac{20 (n_l + n_c)^2 - 29 n_b n_c}{9}
\biggr] \frac{L^2}{8}\\
&&{} + \biggl[
- 129 C_F \left( C_F - \frac{C_A}{2} \right) - \frac{11413}{54} C_A^2
- 96 (C_F - C_A)T_F (n_l + n_c) \zeta_3\\
&&\hphantom{{}+\biggl[\biggr.}
+ 4 C_F T_F \left( 23 n_l + \frac{67}{12} n_c - \frac{11}{12} n_b \right)
+ \frac{8}{3} C_A T_F \left( \frac{139}{9} n_l - \frac{47}{4} n_c - 8 n_b \right)\\
&&\hphantom{{}+\biggl[\biggr.}
+ \frac{8}{27} T_F^2 \left( (n_l + n_c) (35 n_l + 124 n_c) + 124 n_b n_c \right)
\biggr] \frac{L}{64}\\
&&{} + \biggl[
\frac{C_F}{4}
\left( B_4 - \frac{\pi^4}{20} + \frac{57}{8} \zeta_3 - \frac{683}{144} \right)
- \frac{C_A}{8}
\left( B_4 - \frac{\pi^4}{10} + \frac{629}{72} \zeta_3 - \frac{16627}{1944} \right)\\
&&\hphantom{{}+\biggl[\biggr.}
+ \frac{T_F}{18} \left( - \bigl( 4 n_l - 7 (n_b + n_c) \bigr) \zeta_3
+ \frac{2654 n_l - 1685 (n_b + n_c)}{432}
\right)
\biggr] T_F (n_b + n_c)\\
&&{} + \biggl[ - 64 L_+(x) + \frac{(1 + x^2) (5 + 22 x^2 + 5 x^4)}{x^3} L_-(x)
- 96 \zeta_3\\
&&\hphantom{{}+\biggl[\biggr.}
- 5 \left( \frac{(1 - x^2)^2}{x^2} (L^2 + 2) - 2 \frac{1 - x^4}{x^2} L \right)
\biggr] \frac{T_F^2 n_b n_c}{96}
+ \mathcal{O}(\varepsilon)\,.
\end{eqnarray*}
At $x=1$ this result reduces to the ordinary decoupling of $n_b+n_c$ flavours
with the same mass~\cite{Chetyrkin:1997un}.

Specifying to QCD leads to (for $\varepsilon=0$)
\begin{eqnarray}
&&\zeta_m(m_c(\bar{m}_b),\bar{m}_b) = 1
- 2 L \frac{\alpha_s^{(n_f)}(\bar{m}_b)}{\pi}
\nonumber\\
&&{} + \left[
- \left( n_l - \frac{43}{2} \right) \frac{L^2}{3}
+ \left( 5 n_l - \frac{293}{2} \right) \frac{L}{18}
+ \frac{89}{216}
\right] \left(\frac{\alpha_s^{(n_f)}(\bar{m}_b)}{\pi}\right)^2
\nonumber\\
&&{} + \biggl[
- 2 \left( n_l^2 - 40 n_l + \frac{1589}{4} \right) \frac{L^3}{27}
+ \left( \frac{5}{3} n_l^2 - \frac{679}{6} n_l + \frac{2497}{2} \right) \frac{L^2}{18}
\nonumber\\
&&\hphantom{{}+\biggl[\biggr.}
+ \left( 5 \zeta_3 (n_l + 1)
+ \frac{1}{72} \left( \frac{35}{3} n_l^2 + 607 n_l - \frac{103771}{12} \right)
\right) \frac{L}{3}
- \frac{2}{9} L_+(x)
\nonumber\\
&&\hphantom{{}+\biggl[\biggr.}
+ \frac{(1 + x^2) (5 + 22 x^2 + 5 x^4)}{288 x^3} L_-(x)
- \frac{5}{288} \left( \frac{(1 - x^2)^2}{x^2} (L^2 + 2) - 2 \frac{1 - x^4}{x^2} L \right)
\nonumber\\
&&\hphantom{{}+\biggl[\biggr.}
- \frac{1}{18} \biggl( B_4 - \frac{\pi^4}{2} + \frac{8}{3} \zeta_3 n_l - \frac{439}{24} \zeta_3
- \frac{1327}{324} n_l - \frac{21923}{648} \biggr)
\biggr] \left(\frac{\alpha_s^{(n_f)}(\bar{m}_b)}{\pi}\right)^3
+ \cdots\,,
\nonumber\\
\label{m:su3}
\end{eqnarray}
where for $x\ll1$ the coefficient of $(\alpha_s/\pi)^3$ takes the form
\begin{eqnarray*}
&&- 2 \left( n_l^2 - 40 n_l + \frac{1591}{4} \right) \frac{L^3}{27}
+ \left( 5 n_l^2 - \frac{679}{2} n_l + \frac{15011}{4} \right) \frac{L^2}{54}\\
&&{} \left[ 5 \zeta_3 (n_l + 1)
+ \frac{1}{72} \left( \frac{35}{3} n_l^2 + 607 n_l - \frac{104267}{12} \right)
\right] \frac{L}{3}\\
&&{} - \frac{1}{18} \left( B_4 - \frac{\pi^4}{2}
+ \frac{8}{3} \zeta_3 n_l + \frac{439}{24} \zeta_3
- \frac{1327}{324} n_l - \frac{24935}{648} \right)\\
&&{} - \left( 2 L - \frac{47}{30} \right) \frac{x^2}{15}
+ \mathcal{O}(x^4)\,.
\end{eqnarray*}

\section{Decoupling for the fields}
\label{S:fields}

\subsection{Gluon field and the gauge parameter}
\label{S:a}

Decoupling of the gluon field and the gauge fixing parameter
are given by the same quantity $\zeta_A^0$~(cf. (\ref{Intro:zetafields0})):
\begin{equation}
a_0^{(n_l)} = a_0^{(n_f)} \zeta_A^0(\alpha_{s0}^{(n_f)},a_0^{(n_f)},m_{b0},m_{c0})\,.
\label{a:bare}
\end{equation}
In a first step we replace the bare quantities in the right-hand side
via the renormalized ones using Eqs.~(\ref{as:MSbar}),~(\ref{as:massren}),
and~\cite{Larin:1993tp,Chetyrkin:2004mf,Czakon:2004bu}
\begin{equation}
a_0^{(n_f)} = Z_A^{(n_f)}\left(\alpha_s^{(n_f)}(\mu),a^{(n_f)}(\mu)\right) a^{(n_f)}(\mu)\,,
\label{a:rena}
\end{equation}
and thus we express $a_0^{(n_l)}$ via the $n_f$-flavour renormalized quantities.
In a next step we can find $a^{(n_l)}(\mu')$ in terms of $a_0^{(n_l)}$
by solving the equation
\begin{equation}
a_0^{(n_l)} = Z_A^{(n_l)}\left(\alpha_s^{(n_l)}(\mu'),a^{(n_l)}(\mu')\right) a^{(n_l)}(\mu')
\label{a:renl}
\end{equation}
iteratively.
The result reads
\begin{equation}
\zeta_A(m_c(\bar{m}_b),\bar{m}_b) = 1
+ d^A_1 \frac{\alpha_s^{(n_f)}(\bar{m}_b)}{\pi}
+ d^A_2 \left(\frac{\alpha_s^{(n_f)}(\bar{m}_b)}{\pi}\right)^2
+ d^A_3 \left(\frac{\alpha_s^{(n_f)}(\bar{m}_b)}{\pi}\right)^3 + \cdots
\,,
\label{s:renorm}
\end{equation}
where
\begin{eqnarray*}
d^A_1 &=& - \frac{C_A (3 a - 13) + 8 T_F (n_l + n_c)}{12} L\\
&&{} + \left\{
\left[ C_A (3 a - 13) + 8 T_F (n_l + n_c) \right] L^2
+ T_F (n_b + n_c) \frac{\pi^2}{3}
\right\} \frac{\varepsilon}{12}\\
&&{} - \left\{
\left[ C_A (3 a - 13) + 8 T_F (n_l + n_c) \right] L^3
+ T_F n_c \pi^2 L
+ 2 T_F (n_b + n_c) \zeta_3
\right\} \frac{\varepsilon^2}{18} + \mathcal{O}(\varepsilon^3)\,,\\
d^A_2 &=& C_A \frac{2 a + 3}{96}
\left[ C_A (3 a - 13) + 8 T_F (n_l+n_c) \right] L^2\\
&&{} - \left[ C_A^2 \frac{2 a^2 + 11 a - 59}{64}
+ C_F T_F \frac{n_l-n_c}{2} + \frac{5}{8} C_A T_F (n_l+n_c)\right] L\\
&&{} + \frac{13}{192} (4 C_F - C_A) T_F (n_b+n_c)\\
&&{} + \biggl\{
- C_A \frac{2 a + 3}{48} \left[ C_A (3 a - 13) + 8 T_F (n_l + n_c) \right] L^3\\
&&\hphantom{{}+\biggl\{\biggr.}
+ \left[ C_A^2 \frac{2 a^2 + 11 a - 59}{32} + C_F T_F (n_l - 2 n_c)
+ \frac{5}{4} C_A T_F (n_l + n_c) \right] L^2\\
&&\hphantom{{}+\biggl\{\biggr.}
- T_F \left[ 13 C_F n_c
+ C_A \frac{\pi^2 \bigl( n_c (a + 3) + n_b a \bigr) - 39 n_c}{12} \right]
\frac{L}{12}\\
&&\hphantom{{}+\biggl\{\biggr.}
- \left[C_F (2 \pi^2 + 35) - \frac{C_A}{2} \left( 5 \pi^2 + \frac{169}{6} \right) \right]
\frac{T_F (n_b + n_c)}{96}
\biggr\} \varepsilon + \mathcal{O}(\varepsilon^2)\,,\\
d^A_3 &=& \frac{C_A}{18} \biggl[
- C_A^2 \frac{(3 a - 13) (6 a^2 + 18 a + 31)}{64}
- C_A T_F (n_l + n_c) \frac{6 a^2 + 15 a + 44}{8}\\
&&\hphantom{C_A\biggl[\biggr.}
+ T_F^2 \bigl( (n_l + n_c)^2 + n_b n_c \bigr)
\biggr] L^3\\
&&{} + \biggl[
\frac{C_A^3}{128} \left( \frac{5}{2} a^3 + \frac{29}{3} a^2 -  17 a - \frac{3361}{18} \right)
+ C_F C_A T_F \frac{6 a (n_l - n_c) + 31 n_l - 49 n_c}{48}\\
&&\hphantom{C_A\biggl[\biggr.}
+ \frac{C_A^2 T_F (n_l + n_c)}{16} \left( \frac{a^2}{3} + 3 a + \frac{401}{18} \right)
- \frac{C_F T_F^2}{6} \left( n_l^2 - n_c^2 + \frac{11}{16} n_b n_c \right)\\
&&\hphantom{C_A\biggl[\biggr.}
- \frac{C_A T_F^2}{18} \left( 5 (n_l + n_c)^2 + \frac{73}{16} n_b n_c \right)
\biggr] L^2\\
&&{} + \biggl[
- \frac{C_A^3}{1024} \left( 6 \zeta_3 (a+1) (a+3)
+ 7 a^3 + 33 a^2 + 167 a - \frac{9965}{9} \right)
+ C_F^2 T_F \frac{n_l - 9 n_c}{16}\\
&&\hphantom{{}+\biggl[\biggr.}
- \frac{C_F C_A T_F}{4} \left( 3 \zeta_3 (n_l + n_c) + \frac{13}{48} a (n_b + n_c)
+ \frac{1}{36} \left( \frac{5}{4} n_l - 227 n_c \right) \right)\\
&&\hphantom{{}+\biggl[\biggr.}
+ \frac{C_A^2 T_F}{16} \biggl( 9 \zeta_3 (n_l + n_c)
+ a \left( n_l + \frac{61}{48} n_c - \frac{25}{72} n_b \right)\\
&&\hphantom{{}+\biggl[{}+\frac{C_A^2T_F}{16}\biggl(\biggr.\biggr.}
- \frac{1}{36} \left( 911 n_l + \frac{3241}{4} n_c - \frac{1157}{12} n_b \right)
\biggr)\\
&&\hphantom{{}+\biggl[\biggr.}
+ C_F T_F^2 \frac{(n_l + n_c) (11 n_l + 4 n_c) + 4 n_b n_c}{72}\\
&&\hphantom{{}+\biggl[\biggr.}
+ \frac{C_A T_F^2}{32}
\left( \frac{(n_l + n_c) (76 n_l + 63 n_c)}{9}
+ n_b \left( 7 n_c - \frac{178}{54} n_l \right) \right)
\biggr] L\\
&&{} + \biggl[
- \frac{C_F^2}{12} \left( \frac{95}{2} \zeta_3 - \frac{97}{3} \right)
+ C_F C_A \left( B_4 - \frac{\pi^4}{20} + \frac{1957}{96} \zeta_3 - \frac{36979}{2592} \right)\\
&&\hphantom{{}+\biggl[\biggr.}
- \frac{C_A^2}{2}
\left( B_4 - \frac{3 \pi^4}{40} + \frac{\zeta_3 a}{3} + \frac{1709}{288} \zeta_3
- \frac{677}{432} a + \frac{22063}{3888} \right)\\
&&\hphantom{{}+\biggl[\biggr.}
+ \frac{164}{81} C_F T_F n_l
+ C_F T_F (n_b + n_c) \left( \frac{7}{8} \zeta_3 - \frac{103}{162} \right)\\
&&\hphantom{{}+\biggl[\biggr.}
- \frac{C_A T_F n_l}{9} \left( 8 \zeta_3 - \frac{665}{54} \right)
+ \frac{C_A T_F (n_b + n_c)}{18} \left( \frac{287}{8} \zeta_3 - \frac{605}{27} \right)
\biggr] \frac{T_F (n_b + n_c)}{8}\\
&&{} + T_F^2 n_b n_c \biggl[ - \frac{C_A}{3} L_+(x)
+ \frac{1+x^2}{32 x^3} \left( C_F \frac{5 - 2 x^2 + 5 x^4}{4}
+ C_A \frac{4 + 11 x^2 + 4 x^4}{3} \right) L_-(x)\\
&&\hphantom{{}+T_F^2n_bn_c\biggl[\biggr.}
- \frac{14 C_F + 39 C_A}{64} \zeta_3\\
&&\hphantom{{}+T_F^2n_bn_c\biggl[\biggr.}
- \left( \frac{5}{16} C_F + \frac{C_A}{3} \right)
\left( \frac{(1-x^2)^2}{8 x^2} \left( L^2 + 2 \right)
- \frac{1-x^4}{4 x^2} L \right) \biggr]
+ \mathcal{O}(\varepsilon)\,,
\end{eqnarray*}
with $a\equiv a^{(n_f)}(\bar{m}_b)$.
The easiest way to express $a^{(n_f)}(\bar{m}_b)$ via $a^{(n_l)}(m_c(\bar{m}_b))$
is to re-express $\alpha_s^{(n_f)}(\bar{m}_b)$ via $\alpha_s^{(n_l)}(m_c(\bar{m}_b))$
in the right-hand side of the equation
$a^{(n_l)}(m_c(\bar{m}_b)) = a^{(n_f)}(\bar{m}_b) \zeta_A(\bar{m}_b,m_c(\bar{m}_b))$
and then solve it for $a^{(n_f)}(\bar{m}_b)$ iteratively.

\subsection{Light-quark fields}
\label{S:q}

The bare decoupling coefficient $\zeta_q^0$ of Eq.~(\ref{Intro:zetafields0})
is determined by $\Sigma_V(0)$ (cf. Eq.~(\ref{Calc:Sigma})).
The renormalized version $\zeta_q$~(\ref{Intro:zetaren}) can be obtained
(see Refs.~\cite{Larin:1993tp,Chetyrkin:1999pq,Czakon:2004bu} for the
three-loop wave function renormalization constant)
by re-expressing $\alpha_s^{(n_l)}$ and $a^{(n_l)}$ in the denominator
via the $n_f$-flavour quantities (see Sects.~\ref{S:as} and \ref{S:a};
note that positive powers of $\varepsilon$ should be kept).
The result can be cast in the form
\begin{equation}
\zeta_q(m_c(\bar{m}_b),\bar{m}_b) = 1
+ d^q_1 C_F \frac{\alpha_s^{(n_f)}(\bar{m}_b)}{\pi}
+ d^q_2 C_F \left(\frac{\alpha_s^{(n_f)}(\bar{m}_b)}{\pi}\right)^2
+ d^q_3 C_F \left(\frac{\alpha_s^{(n_f)}(\bar{m}_b)}{\pi}\right)^3 + \cdots
\,,
\label{q:renorm}
\end{equation}
where 
\begin{eqnarray*}
d^q_1 &=& - \frac{a}{2} L \left( 1 - L \varepsilon + \frac{2}{3} L^2 \varepsilon^2
+ \mathcal{O}(\varepsilon^3) \right)\,,\\
d^q_2 &=& \frac{a}{16} \left[ 2 C_F a + C_A (a + 3) \right] L^2
+ \left( 6 C_F - C_A (a^2 + 8 a + 25) + 8 T_F (n_l + n_c) \right) \frac{L}{32}\\
&&{} + \frac{5}{96} T_F (n_b + n_c)\\
&&{} - \biggl[ a \left[ 2 C_F a + C_A (a + 3) \right] L^3
+ \left( 6 C_F - C_A (a^2 + 8 a + 25) + 8 T_F (n_l + n_c) \right) \frac{L^2}{2}\\
&&\hphantom{{}-\biggl[\biggr.}
+ \frac{5}{3} T_F n_c L
+ \frac{T_F (n_b + n_c)}{12} \left( \pi^2 + \frac{89}{6} \right)
\biggr] \frac{\varepsilon}{8} + \mathcal{O}(\varepsilon^2)\,,\\
d^q_3 &=& \frac{a}{8} \biggl[
- C_F^2 \frac{a^2}{6} - C_F C_A \frac{a (a+3)}{4} - C_A^2 \frac{2 a^2 + 9 a + 31}{24}
+ C_A T_F \frac{n_l + n_c}{3} \biggr] L^3\\
&&{} + \biggl[
- \frac{3}{32} C_F^2 a
+ C_F C_A \frac{a^3 + 8 a^2 + 25 a - 22}{64}
+ \frac{C_A^2}{64}
\left( a^3 + \frac{25}{4} a^2 + \frac{343}{12} a + \frac{275}{3} \right)\\
&&\hphantom{{}+\biggl[\biggr.}
- T_F \frac{n_l + n_c}{8} \left( C_F (a-1) + C_A \frac{13 a + 94}{12} \right)
+ T_F^2 \frac{(n_l + n_c)^2}{6}
\biggr] L^2\\
&&{} + \biggl[ - \frac{3}{64} C_F^2
- \frac{C_F C_A}{8} \left( 3 \zeta_3 -\frac{143}{16} \right)\\
&&\hphantom{{}+\biggl[\biggr.}
- \frac{C_A^2}{512} \left( 6 \zeta_3 (a^2 + 2 a - 23)
+ 5 a^3 + \frac{39}{2} a^2 + \frac{263}{2} a + \frac{9155}{9} \right)\\
&&\hphantom{{}+\biggl[\biggr.}
- \frac{C_F T_F}{32} \left( \frac{5}{6} (n_b + n_c) a - 3 (n_l + 5 n_c) \right)\\
&&\hphantom{{}+\biggl[\biggr.}
+ \frac{C_A T_F}{288} \left( \frac{153 (n_l+n_c) - 89 n_b}{4} a + 287 n_l + 232 n_c \right)
- \frac{5}{72} T_F^2 n_l (n_l + n_c)
\biggr] L\\
&&{} + \biggl[
- C_F \left( 3 \zeta_3 + \frac{155}{48} \right)
- C_A \left( \zeta_3 (a-3)
- \frac{1}{72} \left( \frac{2387}{8} a + \frac{1187}{3} \right) \right)\\
&&\hphantom{{}+\biggl[\biggr.}
+ \frac{35}{2592} T_F (2 n_l + n_b + n_c)
\biggr] \frac{T_F (n_b + n_c)}{24} + \mathcal{O}(\varepsilon)\,.
\end{eqnarray*}
Note that the power corrections in $x$ drop out in the sum of all diagrams.
For $x=1$ this result reduces to the ordinary decoupling of $n_b+n_c$ flavours
with the same mass~\cite{Chetyrkin:1997un}
(see Ref.~\cite{Grozin:2006xm} for an expression in terms of $C_A$ and $C_F$).

\subsection{Ghost field}
\label{S:c}

The bare decoupling coefficient $\zeta_c^0$ in Eq.~(\ref{Intro:zetafields0})
is determined by $\Pi_c(0)$ as given in Eq.~(\ref{Calc:Ghost}).
The renormalized decoupling constant $\zeta_c$ of Eq.~(\ref{Intro:zetaren}) is
given by (see Refs.~\cite{Chetyrkin:2004mf,Czakon:2004bu} for the
corresponding renormalization constant)
\begin{equation}
\zeta_c(m_c(\bar{m}_b),\bar{m}_b) = 1
+ d^c_1 C_A \frac{\alpha_s^{(n_f)}(\bar{m}_b)}{\pi}
+ d^c_2 C_A \left(\frac{\alpha_s^{(n_f)}(\bar{m}_b)}{\pi}\right)^2
+ d^c_3 C_A \left(\frac{\alpha_s^{(n_f)}(\bar{m}_b)}{\pi}\right)^3 + \cdots
\,,
\label{c:renorm}
\end{equation}
where 
\begin{eqnarray*}
d^c_1 &=& - \frac{a-3}{8} L
\left( 1 - L \varepsilon + \frac{2}{3} L^2 \varepsilon^2
+ \mathcal{O}(\varepsilon^3) \right)\,,\\
d^c_2 &=&
\left[ C_A \frac{3 a^2 - 35}{16} + T_F (n_l + n_c) \right] \frac{L^2}{8}
+ \left[ C_A \frac{3 a + 95}{8} - 5 T_F (n_l + n_c) \right] \frac{L}{48}\\
&&{} - \frac{89}{1152} T_F (n_b + n_c)\\
&&{} + \biggl\{
- \left[ C_A \frac{3 a^2 - 35}{16} + T_F (n_l + n_c) \right] \frac{L^3}{4}
- \left[ C_A \frac{3 a + 95}{8} - 5 T_F (n_l + n_c) \right] \frac{L^2}{24}\\
&&\hphantom{{}+\biggl\{\biggr.}
- T_F \frac{3 \pi^2 n_b - 89 n_c}{288} L
+ \frac{T_F (n_b + n_c)}{1152} \left( 5 \pi^2 + \frac{869}{6} \right)
\biggr\} \varepsilon + \mathcal{O}(\varepsilon^2)\,,\\
d^c_3 &=& \biggl[
- \frac{C_A^2}{256} \left( 5 a^3 + 9 a^2 - \frac{35}{3} a - \frac{2765}{9} \right)
- C_A T_F (n_l + n_c) \frac{3 a + 149}{144}\\
&&\hphantom{\biggl[\biggr.}
+ \frac{T_F^2}{9} \bigl( 2 (n_l + n_c)^2 - n_b n_c \bigr)
\biggr] \frac{L^3}{4}\\
&&{} + \biggl[
\frac{C_A^2}{16} \left( a^3 + \frac{9}{2} a^2 - \frac{11}{3} a - \frac{5773}{18} \right)
+ \left( 3 C_F + C_A \frac{3 a + 545}{36} \right) T_F (n_l + n_c)\\
&&\hphantom{{}+\biggl[\biggr.}
- \frac{T_F^2}{9} \bigl( 20 (n_l + n_c)^2 - 29 n_b n_c \bigr)
\biggr] \frac{L^2}{32}\\
&&{} + \biggl[
\frac{C_A^2}{128} \left( 3 \zeta_3 (a+1) (a+3)
- \frac{3}{2} a^3 - 3 a^2 - 17 a + \frac{15817}{54} \right)\\
&&\hphantom{{}+\biggl[\biggr.}
+ C_F T_F \left( 3 \zeta_3 (n_l + n_c)
- \frac{45 n_l + 25 n_c + 13 n_b}{16} \right)\\
&&\hphantom{{}+\biggl[\biggr.}
+ \frac{C_A T_F}{32} \left( - 72 \zeta_3 (n_l + n_c)
+ \frac{252 n_l + 341 n_c - 89 n_b}{36} a
- \frac{194}{27} n_l + \frac{695 n_c + 167 n_b}{12}
\right)\\
&&\hphantom{{}+\biggl[\biggr.}
- \frac{T_F^2}{27} \left( \frac{(n_l + n_c) (35 n_l + 124 n_c)}{4} + 31 n_b n_c \right)
\biggr] \frac{L}{8}\\
&&{} + \biggl[
- \frac{C_F}{2} \left( B_4 - \frac{\pi^4}{20} + \frac{57}{8} \zeta_3 - \frac{481}{96} \right)\\
&&\hphantom{{}+\biggl[\biggr.}
+ \frac{C_A}{4} \left( B_4 - \frac{3 \pi^4}{40} - \frac{\zeta_3 a}{3} + \frac{431}{72} \zeta_3
+ \frac{685}{864} a - \frac{5989}{1944} \right)\\
&&\hphantom{{}+\biggl[\biggr.}
+ \frac{4}{9} T_F n_l \left( \zeta_3 - \frac{1327}{864} \right)
- \frac{T_F (n_b + n_c)}{9} \left( 7 \zeta_3 - \frac{1685}{432} \right)
\biggr] \frac{T_F (n_b + n_c)}{8}\\
&&{} + \frac{T_F^2 n_b n_c}{6} \biggl[ L_+
- \frac{(1 + x^2) (5 + 22 x^2 + 5 x^4)}{64 x^3} L_-
+ \frac{3}{2} \zeta_3\\
&&\hphantom{{}+\frac{T_F^2n_bn_c\biggl[\biggr.}{}}
+ \frac{5}{64} \left( \frac{(1 - x^2)^2}{x^2} (L^2 + 2) - 2 \frac{1 - x^4}{x^2} L \right)
\biggr] + \mathcal{O}(\varepsilon)\,.
\end{eqnarray*}


\section{Phenomenological applications}
\label{S:Phen}

In this section we study the numerical consequences of the decoupling
relations computed in the previous sections.
For convenience we use in this Section the decoupling relations in
terms of on-shell heavy quark masses (see Appendix~\ref{S:OS} and the
{\tt Mathematica} file which can be downloaded from~\cite{progdata})
which we denote by $M_c$ and $M_b$.

\subsection{$\alpha_s^{(5)}(M_Z)$ from $\alpha_s^{(3)}(M_\tau)$}

Let us in a first step check the dependence on the decoupling scales 
which should become weaker after including higher order perturbative
corrections. We consider the relation bet\-ween $\alpha_s^{(3)}(M_\tau)$ and
$\alpha_s^{(5)}(M_Z)$. $\alpha_s^{(3)}(M_\tau)$ has been extracted from
experimental data using perturbative results up to order
$\alpha_s^4$~\cite{Baikov:2008jh}. Thus it is mandatory to perform the
transition from the low 
to the high scale with the highest possible precision. In the following we
compare the conventional approach with the single-step decoupling up to
three-loop order. 

For our analysis we use for convenience the decoupling constants expressed in
terms of on-shell quark masses. In this way the mass values are fixed and they
are not affected by the running from $M_\tau$
to $M_Z$. In our analysis we use $M_c=1.65$~GeV and
$M_b=4.7$~GeV. Furthermore,
$\alpha_s^{(3)}(M_\tau) = 0.332$~\cite{Baikov:2008jh} is used as starting
value of our analysis.

\FIGURE[t]{
  \begin{tabular}{cc}
  \leavevmode
  \epsfxsize=.45\textwidth
  \epsffile{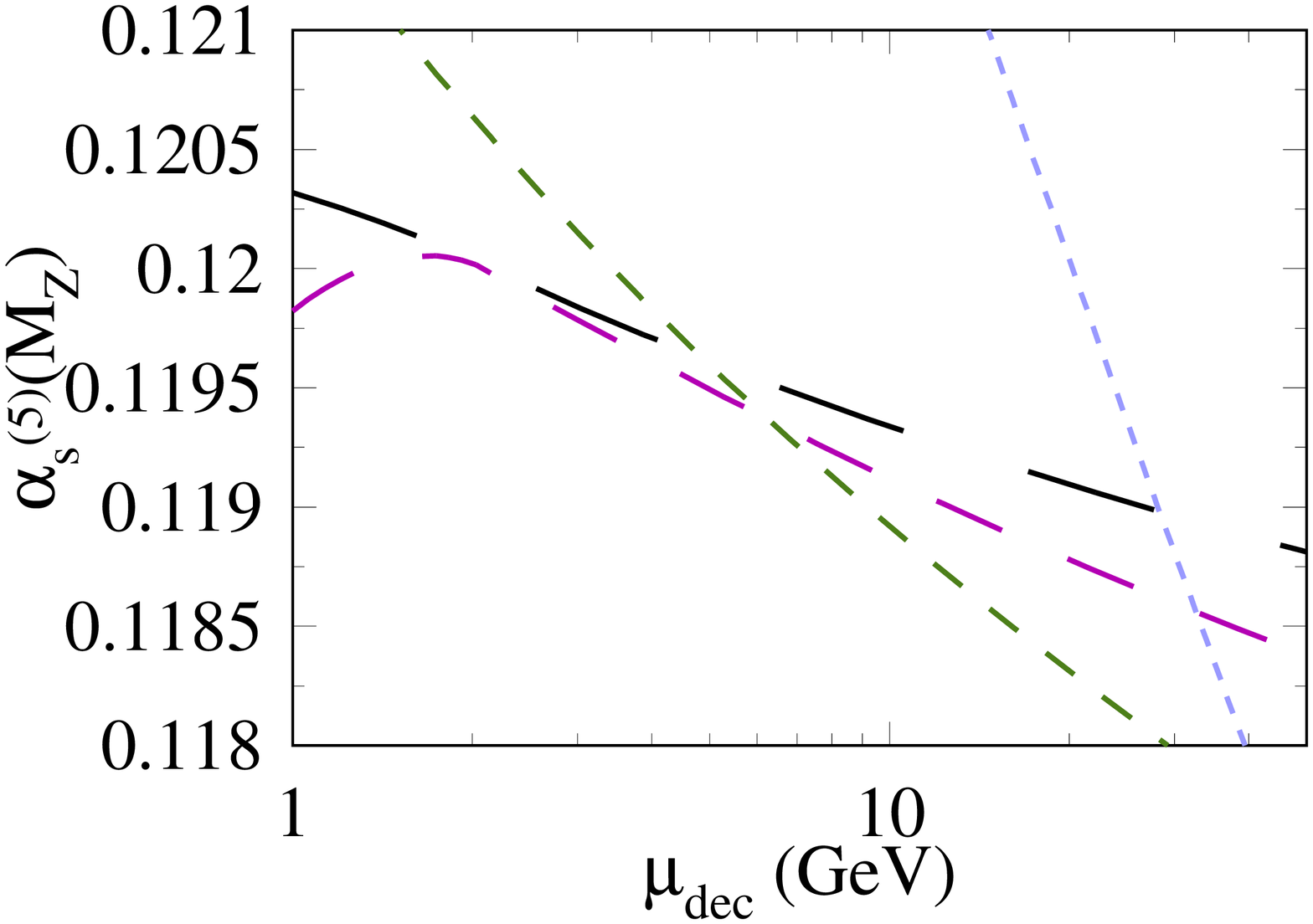}
  &
  \leavevmode
  \epsfxsize=.45\textwidth
  \epsffile{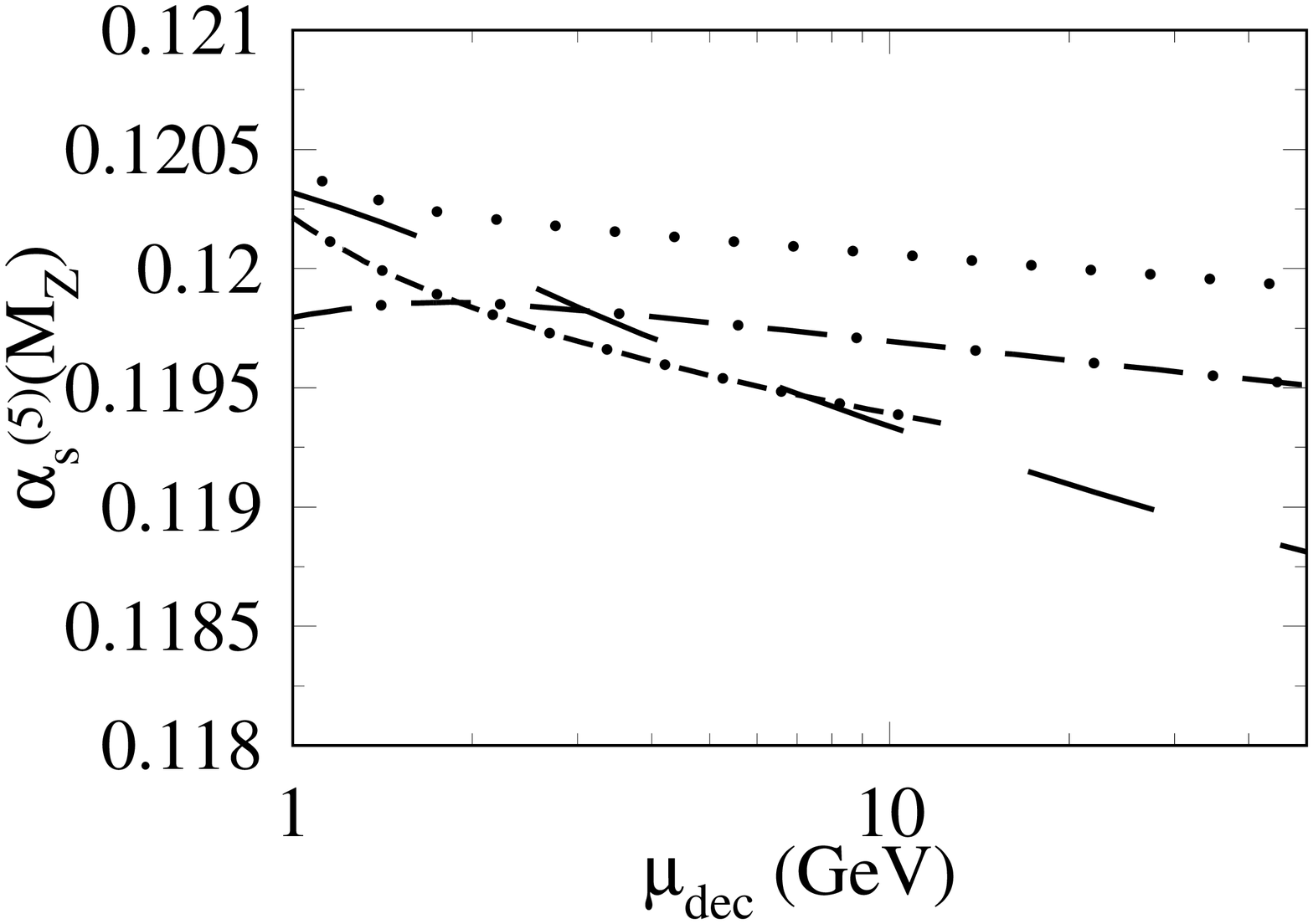}
  \\ (a) & (b)
  \end{tabular}
  \caption{\label{fig::asMz}
    $\alpha_s^{(5)}(M_Z)$ as obtained from $\alpha_s^{(3)}(M_\tau)$ 
    as a function $\mu_{\rm dec}$.
    The dashed lines (long dashes include higher order perturbative results) 
    correspond to the single-step approach and the dash-dotted
    curves 
    (short dashes: $\mu_{{\rm dec},c}=\mu_{{\rm dec}}$, 
    long dashes: $\mu_{{\rm dec},b}=\mu_{{\rm dec}}$)
    are obtained in the conventional analysis using four-loop running
    and three-loop decoupling relations. 
    The dotted line results from a five-loop analysis of the two-step
    (see text for details).
    }
}

In Fig.~\ref{fig::asMz}(a) we show $\alpha_s^{(5)}(M_Z)$ as a function of
$\mu_{\rm dec}$, the scale where the $c$ and $b$ quarks are simultaneously
integrated out.
In a first step $\alpha_s^{(3)}(M_\tau)$ is evolved to
$\alpha_s^{(3)}(\mu_{\rm dec})$ using the $N$-loop renormalization group
equations. Afterwards the $(N-1)$-loop decoupling relation is applied and
finally $N$-loop running is employed in order to arrive at
$\alpha_s^{(5)}(M_Z)$. One observes a strong dependence on $\mu_{\rm dec}$ for
$N=1$ (short-dashed line) which becomes rapidly weaker when increasing $N$ 
leading to a reasonably
flat curve for $N=4$ (longer dashes correspond to larger values
of $N$). 

\subsection{Comparison of one- and two-step decoupling approach}

In the step-by-step decoupling  
approach we have two decoupling scales $\mu_{{\rm dec},c}$ and 
$\mu_{{\rm dec},b}$ which can be chosen independently. First
we choose\footnote{It has
  been argumented in Refs.~\cite{Kuhn:2007vp} 
  that in the case of charm the scale $\mu=m_c$ is too small leading to a
  value of $\alpha_s$ which is too large. Thus $m_c(3~\mbox{GeV})$ has been
  proposed as reference value.}
$\mu_{{\rm dec},c}=3$~GeV
and identify $\mu_{{\rm dec},b}$ with $\mu_{{\rm dec}}$. The result for $N=4$
is shown in Fig.~\ref{fig::asMz}(b) together with the four-loop curve from
Fig.~\ref{fig::asMz}(a) as dash-dotted line (long dashes). One
observes a significantly flatter behaviour as for the one-step decoupling
which can be explained by the occurrence of $\log(\mu^2/M_c^2)$ terms in the
one-step formula which might become large for large values of $\mu=\mu_{{\rm dec}}$.
Alternatively it is also possible to study the dependence on 
$\mu_{{\rm dec},c}$, i.e., identify $\mu_{{\rm dec},c}$ with $\mu_{{\rm
    dec}}$, set $\mu_{{\rm dec},b}=10$~GeV
and compare to the one-step decoupling. The results are also shown in
Fig.~\ref{fig::asMz}(b) as dash-dotted line (short dashes) where only
values $\mu_{{\rm dec}}\le 10$~\mbox{GeV} are considered.

For comparison we show in Fig.~\ref{fig::asMz}(b) also the result of the
two-step five-loop analysis as dotted line 
where the four-loop decoupling relation is taken from
Refs.~\cite{Schroder:2005hy,Chetyrkin:2005ia}. 
The (unknown) five-loop coefficient of the $\beta$ function,
$\beta_4$, is set to zero.\footnote{For $\beta_4>0$ the dotted curve in
  Fig.~\ref{fig::asMz}(b) moves towards the four-loop curve.}
If one restricts to scales $\mu_{\rm dec}$ between 2~GeV and 10~GeV it seems
that the four-loop decoupling constant is numerically more relevant than
the
power-suppressed terms included by construction in the one-step decoupling
procedure. Thus, from these considerations one tends to prefer the two-step
decoupling over the one-step approach as it seems that the resummation of 
$\log(\mu^2/M_{c,b}^2)$ is more important than the inclusion of
power-suppressed corrections. 

Let us in a next step restrict ourselves to decoupling scales which are of the
order of the respective quark masses. 
In Tab.~\ref{tab::mub} we compare the value for $\alpha_s^{(5)}(M_Z)$
as obtained from the one- and two-step decoupling where two variants of the
former are used: $\zeta_{\alpha_s}$ which directly
relates $\alpha_s^{(3)}(\mu_c)$ and $\alpha_s^{(5)}(\mu_b)$ as 
given in Eq.~(\ref{Intro:zetaren}) with $\mu^\prime=\mu_c$ and $\mu=\mu_b$
($\zeta_{\alpha_s}(\mu_c,\mu_b)$; see also~\cite{progdata}) 
and the version with only one decoupling scale
where $\mu^\prime=\mu$ has been set ($\zeta_{\alpha_s}(\mu)$).
We thus define two deviations
\begin{eqnarray}
  \delta\alpha_s^{(a)} &=&
  \alpha_s^{(5)}(M_Z)\Big|_{\zeta_{\alpha_s}(\mu_c,\mu_b)}
  - \alpha_s^{(5)}(M_Z)\Big|_{\mbox{\scriptsize 2-step}}
  \,,\nonumber\\
  \delta\alpha_s^{(b)} &=&
  \alpha_s^{(5)}(M_Z)\Big|_{\zeta_{\alpha_s}(\mu)}
  - \alpha_s^{(5)}(M_Z)\Big|_{\mbox{\scriptsize 2-step}}
  \,,
  \label{eq::deltaalpha}
\end{eqnarray}
where the scale $\mu$ in the second equation is either identified with $\mu_c$
(right part of Tab.~\ref{tab::mub}) or $\mu_b$ (left part), respectively.

\begin{table}[t]
  {
  \begin{center}
    \begin{tabular}{cc}
      \begin{tabular}{c|r|r|r}
        $\mu_b$ & $\alpha_s^{(5)}(M_Z)$ & $\delta\alpha_s^{(a)}$ &
        $\delta\alpha_s^{(b)}$ \\
        (GeV) &&$\times10^3$& $\times10^3$\\
              &&& $(\mu=\mu_b)$ \\
        \hline
        2&  0.11985& $-0.28$&  0.18\\
        5&  0.11977&   0.23& $-0.16$\\
        7&  0.11974&   0.36& $-0.26$\\
        10& 0.11970&   0.19& $-0.36$    
      \end{tabular}
      &
      \begin{tabular}{c|r|r|r}
        $\mu_c$ & $\alpha_s^{(5)}(M_Z)$ & $\delta\alpha_s^{(a)}$ &
        $\delta\alpha_s^{(b)}$ \\
        (GeV) &&$\times10^3$& $\times10^3$\\
              &&& $(\mu=\mu_c)$ \\
        \hline
        2&  0.11984& $-4.02$& $0.20$\\
        3&  0.11970&   0.19&  $0.14$\\
        4&  0.11961&   0.33&  $0.10$\\
        5&  0.11955&   0.26&  $0.06$\\
      \end{tabular}
    \end{tabular}
    \caption{\label{tab::mub}
      Decoupling scale $\alpha_s^{(5)}(M_Z)$ as
      obtained from the four-loop analysis of the two-step approach, and the
      deviations as defined in the text.
      In the left table $\mu_c=3$~GeV and in the right one
      $\mu_b=10$~GeV has been chosen.}
  \end{center}
  }
\end{table}

It is interesting to note that (except for the choice 
$\mu_c=2$~GeV and $\mu_b=10$~GeV)
the deviations presented in Tab.~\ref{tab::mub} amount to 
about 30\% to 50\% of the uncertainty
of the world average for $\alpha_s(M_Z)$ which is
given by $\delta\alpha_s=0.7\cdot 10^{-3}$~\cite{Nakamura:2010zzi}.

\subsection{Improving the two-step approach by power-suppressed terms}

From the previous considerations it is evident that the resummation of
logarithms of the form $[\alpha_s \log(\mu_c/\mu_b)]^k$, which is automatically
incorporated in the two-step approach, is numerically more important than
power-suppressed terms in $M_c/M_b$. Thus it is natural to use the
two-step approach as default method and add the power-corrections
afterwards.
This is achieved in the following way: In a first step we invert
$\zeta_{\alpha_s}(\mu_c,\mu_b)$ (cf. Eq.~(\ref{Intro:zetaren})) and express it
in terms of $\alpha_s^{(3)}(\mu_c)$ in order to arrive at the equation
$\alpha_s^{(5)}(\mu_b)=\zeta^{-1}_{\alpha_s}(\mu_c,\mu_b)
\alpha_s^{(3)}(\mu_c)$. Now an expansion is performed for
$M_c/M_b\to 0$ to obtain the leading term which is then subtracted from
$\zeta^{-1}_{\alpha_s}(\mu_c,\mu_b)$ since it is part of the
two-step decoupling procedure. The result is independent of $\mu_c$ and
$\mu_b$ and has following series expansion
\begin{eqnarray}
  \delta\zeta_{\alpha_s}^{-1} &=& \left(\frac{\alpha_s^{(3)}(\mu_c)}{\pi}\right)^3
  \left[ 
    \frac{\pi^2}{18}x 
    + \left(-\frac{6661}{18000} - \frac{1409}{21600}L +
      \frac{1}{160}L^2 \right) x^2
    + {\cal O}(x^3)
  \right]
  \nonumber\\
  &\approx& 0.170 \left(\frac{\alpha_s^{(3)}(\mu_c)}{\pi}\right)^3 
  \,,
  \label{eq::deltazeta}
\end{eqnarray}
where the numerical value in the second line
has been obtained with the help of the exact
dependence on $x$. Note that the linear term in $x$ arises from the 
$\overline{\rm MS}$--on-shell quark mass relation.
The quantity $\delta\zeta_{\alpha_s}^{-1}$ is used in order to compute an
additional contribution to $\alpha_s^{(5)}(\mu_b)$ as obtained from the
two-step method:
\begin{eqnarray}
  \delta\alpha_s^{(5)}(\mu_b) &=& \delta\zeta_{\alpha_s}^{-1} \alpha_s^{(3)}(\mu_c)
  \,.
\end{eqnarray}
Inserting numerical values leads to shifts which are at most a few times
$10^{-5}$ and are thus beyond the current level of accuracy.
It is in particular more than an order of magnitude smaller than the four-loop
decoupling term which is shown as dotted curve in Fig.~\ref{fig::asMz}(b).

Note that as far as the strong coupling in Eq.~(\ref{eq::deltazeta}) is
concerned both the number of flavours and the renormalization
scale of $\alpha_s$ are not fixed since power-suppressed terms appear for the
first time at this order. However, the smallness of the contribution is not
affected by the choices made in Eq.~(\ref{eq::deltazeta}).

\subsection{One-step decoupling of the bottom quark with finite charm
    quark mass}

An alternative approach to implement power-suppressed corrections in 
$m_c/m_b$ in the decoupling procedure is as follows: We consider the
step-by-step decoupling and use at the scale $\mu_{\rm dec,c}$ the standard
formalism for the decoupling of the charm quark
as implemented in {\tt RunDec}~\cite{Chetyrkin:2000yt}.
At the scale $\mu_{\rm dec,b}$, however, we consider the matching of five- to
four-flavour QCD where we keep the charm quark massive. This requires a
modification of the formulae in Eqs.~(\ref{Intro:zetafields0})
and~(\ref{Intro:zetaparams0}) to ($n_f^\prime=n_f-1$)
\begin{eqnarray}
&&\zeta_A^0 = \frac{1 + \Pi_A^{(n_f)} (0)}{1 + \Pi_A^{(n_f^\prime)}
    (0)}\,, \quad
  \zeta_c^0 = \frac{1 + \Pi_c^{(n_f)} (0)}{1 + \Pi_c^{(n_f^\prime)}
    (0)}\,, \quad
  \zeta_q^0 = \frac{1 + \Pi_q^{(n_f)} (0)}{1 + \Pi_q^{(n_f^\prime)}
    (0)}\,, \nonumber\\
&&\zeta_m^0 = (\zeta_q^0)^{-1} \frac{1 - \Sigma_S^{(n_f)} (0)}{1
    - \Sigma_S^{(n_f^\prime)} (0)}\,, \quad
  \zeta_{\alpha_s}^0 = (\zeta_c^0)^{-2} (\zeta_A^0)^{-1} \frac{\left(1 +
      \Gamma_{A \bar{c} c}^{(n_f)} \right)^2}{\left(1 + \Gamma_{A
        \bar{c} c}^{(n_f^\prime)} \right)^2}\,,
  \label{eq::2step_mv_c}
\end{eqnarray}
where the $n_f$-flavour quantities contain contributions form massive charm
and bottom quarks. 
They are identical to the one-step decoupling procedure described
above. In the $n_f^\prime$-flavour quantities appearing in the denominators
those diagrams have to be considered which contain a charm quark.
Note that they depend on
the bare parameters of the effective theory 
($\alpha_{s 0}^{(n_f^\prime)}$, $a_0^{(n_f^\prime)}$, $m_{c 0}^{(n_f^\prime)}$)
and thus they
have to be decoupled iteratively in order to express all quantities on
the r.h.s. of the above
equations by the same parameters ($\alpha_{s 0}^{(n_f)}$,
$a_0^{(n_f)}$, $m_{c 0}^{(n_f)}$).
In the standard approach the $n_f^\prime$-flavour quantities vanish since 
only scale-less integrals are involved. 

As a cross check we have verified that we reobtain the analytical result for
the single-step decoupling if we apply the formalism of
Eq.~(\ref{eq::2step_mv_c}) and the subsequent decoupling of the charm quark at
the same scale.

We have incorporated the finite charm quark mass effects in the two-step
decoupling approach (cf. Fig.~\ref{fig::asMz}) and observe small
numerical effects. A minor deviation from the $m_c=0$ curve can only be seen
for decoupling scales of the order of 1~GeV which confirms the conclusions
reached above that the power-suppressed terms are numerically negligible. Thus
we both refrain from 
explicitly presenting numerical results and analytical formulae for the
renormalized decoupling coefficients as obtained from
Eqs.~(\ref{eq::2step_mv_c}).

\subsection{Decoupling effects in the strange quark mass}

In analogy to the strong coupling we study in the following the relation of the
strange quark mass $m_s(\mu)$ defined with three and five active quark
flavours, respectively. 
The numerical analysis follows closely the one for $\alpha_s$: $N$-loop
running is accompanied by $(N-1)$-loop decoupling relations. It is, however,
slightly more involved since besides $m_s(\mu)$ also $\alpha_s(\mu)$ has to be
known for the respective renormalization
scale and number of active flavours. We organized the
calculation in such a way that we simultaneously solve the renormalization group
equations for $m_s(\mu)$ and $\alpha_s(\mu)$ (truncated to the considered
order)  using {\tt Mathematica}.

\FIGURE[t]{
  \leavevmode
  \epsfxsize=\textwidth
  \epsffile{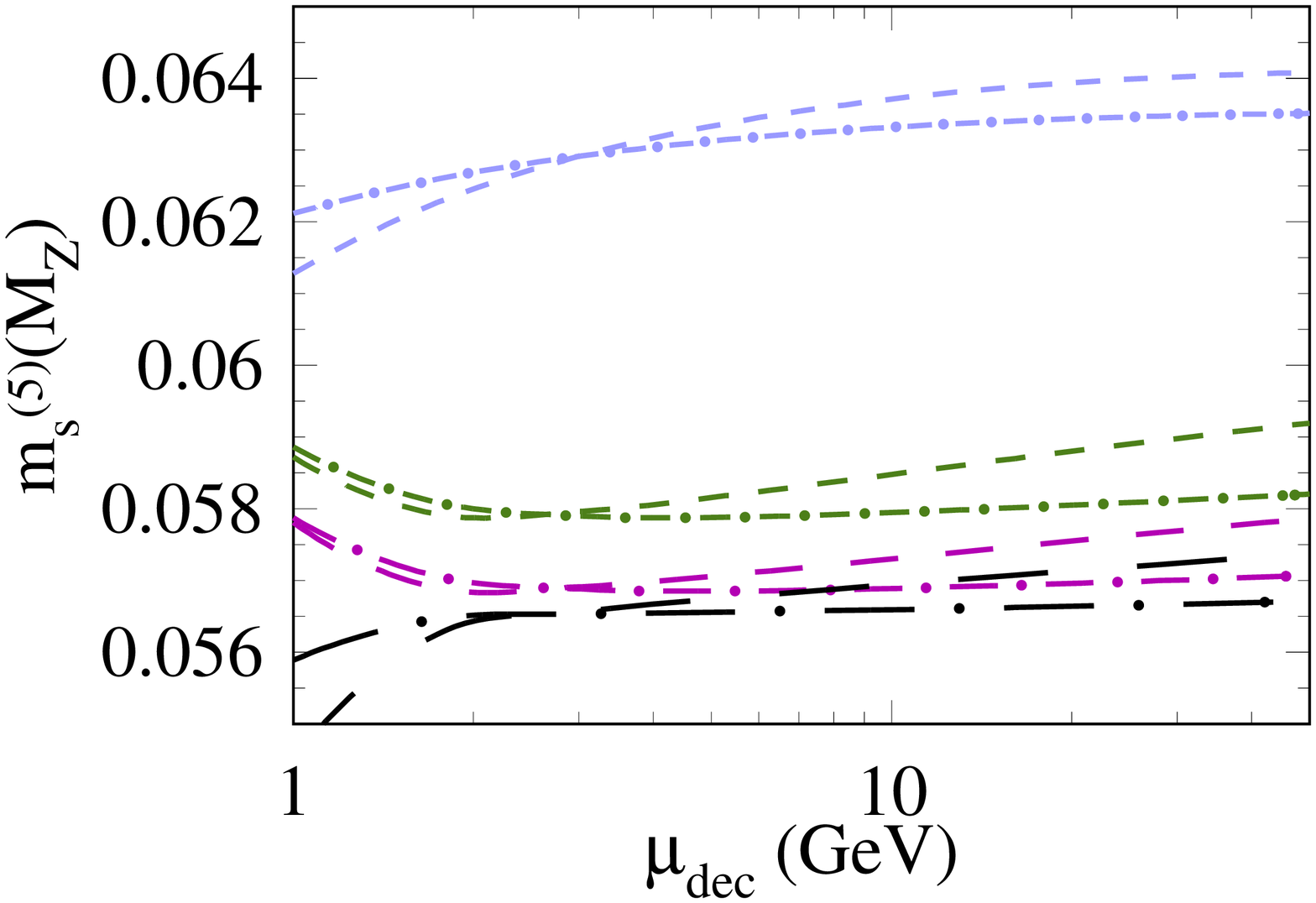}
  \caption{\label{fig::mqMz}
    $m_s^{(5)}(M_Z)$ as a function of $\mu_{\rm dec}$.
    The dashed lines correspond to the single-step approach and the dash-dotted
    curves are obtained in the conventional analysis
    (with $\mu_{{\rm dec},c}=3$~GeV and $\mu_{{\rm dec},b}=\mu_{\rm
      dec}$). Longer dashes correspond to higher loop orders.
    See text for more details.}
}

In Fig.~\ref{fig::mqMz} we show $m_s^{(5)}(M_Z)$ as a function of 
$\mu_{\rm dec}$ and again compare the single-step (dashed lines) to the
two-step (dash-dotted lines) approach. For our numerical
analysis we use in addition to the parameters specified above
$m_s(2~\mbox{GeV})=100$~MeV.
The same conclusion as for $\alpha_s$ can be drawn: The difference between the
two approaches becomes smaller with increasing loop order. At the same time
the prediction for $m_s^{(5)}(M_Z)$ becomes more and more independent of
$\mu_{\rm dec}$. The results again suggest that the power-corrections 
$M_c/M_b$ are
small justifying the application of the two-step decoupling.


\section{Effective coupling of the Higgs boson to gluons}
\label{S:Higgs}

The production and decay of an intermediate-mass Higgs boson can be described
to good accuracy by an effective Lagrange density where the top quark is
integrated out. It contains an effective coupling of the Higgs boson to gluons
given by
\begin{eqnarray}
  {\cal L}_{\rm eff} &=& -\frac{\phi}{v} C_1 {\cal O}_1 
  \,,
  \label{eq::leff}
\end{eqnarray}
with ${\cal O}_1 = G_{\mu\nu} G^{\mu\nu}$.
$C_1$ is the coefficient function containing the remnant
contributions of the top quark, $G^{\mu\nu}$ is the gluon field strength
tensor, $\phi$ denotes the CP-even Higgs boson field and $v$ is the vacuum
expectation value.

The effective Lagrange density in Eq.~(\ref{eq::leff}) can also be used for
theories beyond the Standard Model like supersymmetric models or extensions
with further generations of heavy quarks. In all cases the effect of the heavy
particles is contained in the coefficient function $C_1$.

In Ref.~\cite{Chetyrkin:1997un} a low-energy theorem has been derived which
relates the effective Higgs-gluon coupling $C_1$ to the decoupling constant
for $\alpha_s$. In this Section we apply this theorem to an extension of the
Standard Model containing additional heavy quarks which couple to the Higgs
boson via a top quark-like Yukawa coupling.
Restating Eq.~(39) of Ref.~\cite{Chetyrkin:1997un} in our notation and for the
case of several heavy quarks leads to
\begin{eqnarray}
  C_1 &=& -\frac{1}{2} \sum_{i=1}^{N_h} M_i^2 \frac{{\rm d}}{{\rm d} M_i^2}
  \log\zeta_{\alpha_s}
  \,,
\end{eqnarray}
where $N_h$ is the number of heavy quarks with on-shell masses $M_i$. Using
$\zeta_{\alpha_s}$ from Eq.~(\ref{Intro:zetaren}) (see also~\cite{progdata})
we obtain for $C_1$ the following
result\footnote{Note that up to three-loop order there are only diagrams with 
  at most two different quark flavours. Thus it is possible to obtain the
  result for $C_1$ for $N_h$ heavy quarks.}
\begin{eqnarray}
  C_1 &=& \frac{\alpha_s^{\rm (full)}(\mu)}{\pi} \left( - T_F \frac{N_h}{6} \right) 
  + \left( \frac{\alpha_s^{\rm (full)}(\mu)}{\pi} \right)^2 \left( \frac{C_F T_F}{8}
  - C_A T_F \frac{5}{24} + T_F^2 \frac{\Sigma_h}{18} \right) N_h \nonumber\\
  &&{} + \left( \frac{\alpha_s^{\rm (full)}(\mu)}{\pi} \right)^3 
  \left\{ - C_F^2 T_F \frac{9}{64} N_h
  + C_F C_A T_F \left[ \frac{25}{72} N_h
  + \frac{11}{96} \Sigma_h \right] \right. \nonumber\\
  &&{} + C_F T_F^2 \left[ \frac{5}{96} N_h n_l + \frac{17}{288} N_h^2
  - \Sigma_h \left( \frac{N_h}{8} + \frac{n_l}{12} \right) \right]
  - C_A^2 T_F \left[ \frac{1063}{3456} N_h
  + \frac{7}{96} \Sigma_h \right] \nonumber\\
  &&{} \left. + C_A T_F^2 \left[ \frac{47}{864} n_l - \frac{49}{1728} N_h
  + \frac{5}{24} \Sigma_h \right] N_h
  - T_F^3 \Sigma_h^2 \frac{N_h}{54} \right\}
  \,,
\end{eqnarray}
where $\alpha_s^{\rm (full)}$ is the strong coupling in the full theory with
$n_l+N_h$ active quark flavours and $\Sigma_h = \sum_{i=1}^{N_h} \log(\mu^2/M_i^2)$.
After expressing $\alpha_s^{\rm (full)}$ in terms of $\alpha_s^{(5)}$ and
specifying the colour factors to SU(3) we
reproduce the result of Ref.~\cite{Anastasiou:2010bt} which has been obtained
by an explicit calculation of the Higgs-gluon vertex corrections. 
For $N_h=1$ the result obtained in Ref.~\cite{Chetyrkin:1997un} is reproduced.
It is remarkable that although $\zeta_{\alpha_s}$ contains di- and
tri-logarithms there are only linear logarithms present in $C_1$.


\section{Conclusion}
\label{S:Conc}

The main result of this paper is the computation of a decoupling constant
relating the strong coupling defined with three active flavours to the one in
the five-flavour theory. At three-loop order Feynman diagrams with two mass
scales, the charm and the bottom quark mass, have to be considered. The
corresponding integrals have been evaluated exactly and analytical results have
been presented.
The new results can be used in order to study the effect of power-suppressed
terms in $M_c/M_b$ which are neglected in the conventional
approach~\cite{Chetyrkin:1997un}. Various 
analyses are performed which indicate that the mass corrections present in the
one-step approach are small as
compared to $\log(\mu^2/M_{c,b}^2)$ which are resummed using the conventional
two-step procedure.

Using a well-known low-energy theorem~\cite{Chetyrkin:1997un} we can
use our result for the decoupling 
constant in order to obtain the effective gluon-Higgs boson coupling
for models containing several heavy quarks which couple to the Higgs
boson via the same mechanism as the top quark. 
This constitutes a first independent check of the result presented in
Ref.~\cite{Anastasiou:2010bt} where the matching coefficient has been
obtained by a direct evaluation of the Higgs-gluon-gluon vertex diagrams.


\section*{Acknowledgments}

This work was supported by the Deutsche Forschungsgemeinschaft through
the SFB/TR-9 ``Computational Particle Physics''.
We are grateful to K.\,G.~Chetyrkin for useful discussions.

\appendix



\section{Integral $I(x)$}
\label{S:Ix}

With the help of \texttt{FIRE}~\cite{Smirnov:2008iw} we can express
the integral $I(x)$ as defined in Eq.~(\ref{Calc:Idef}) as a linear
combination of master integrals
\begin{eqnarray}
&&I(x) = I(x^{-1}) =
\frac{1}{(d-1) (d-4) (d-6) (d-8) (d-10)}
\label{Ix:I}\\
&&{}\times\biggl[ \frac{1}{4} \left( c_{10} + c_{11} (x^{-2}+x^2) + c_{12} (x^{-4}+x^4) \right) I_1(x)
\nonumber\\
&&\hphantom{{}\times\biggl[\biggr.}
+ \frac{3}{16} (d-2) (x^{-1}+x) \left( c_{20} + c_{21} (x^{-2}+x^2) \right) I_2(x)
\nonumber\\
&&\hphantom{{}\times\biggl[\biggr.}
- \frac{c_{-1} (x^{2+\varepsilon} + x^{-2-\varepsilon})
+ c_0 (x^\varepsilon + x^{-\varepsilon})
+ c_1 (x^{-2+\varepsilon} + x^{2-\varepsilon})
+ c_2 (x^{-4+\varepsilon} + x^{4-\varepsilon})}%
{(d-2)^2 (d-3) (d-5) (d-7)} \biggr]\,.
\nonumber
\end{eqnarray}
$I_1$ and $I_2$ are master integrals with four massive lines (see Fig.~\ref{F:Master})
which are given by
\FIGURE[t]{\begin{picture}(100,30)
\put(50,15){\makebox(0,0){\includegraphics{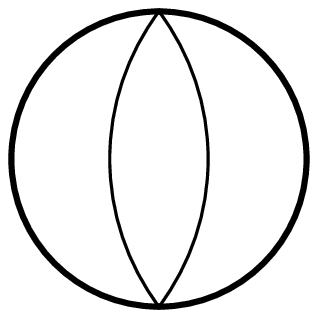}}}
\end{picture}
\caption{Master integral $I_1$ with four massive lines. Thick and
    thin straight lines correspond to
    $b$ and $c$ quarks, respectively. Master integral $I_2$ contains an
    additional numerator.}
\label{F:Master}
}
\begin{eqnarray}
&&I_1(x) = I_1(x^{-1}) =
\frac{(m_b m_c)^{-2+3\varepsilon}}{(i \pi^{d/2})^3 \Gamma^3(\varepsilon)}
\int \frac{d^d k_1\,d^d k_2\,d^d k_3}{D_1 D_2 D_3 D_4}\,,
\nonumber\\
&&I_2(x) = I_2(x^{-1}) =
\frac{(m_b m_c)^{-3+3\varepsilon}}{(i \pi^{d/2})^3 \Gamma^3(\varepsilon)}
\int \frac{N\,d^d k_1\,d^d k_2\,d^d k_3}{D_1 D_2 D_3 D_4}\,,
\nonumber\\
&&D_1 = m_b^2 - k_1^2\,,\quad
D_2 = m_b^2 - k_2^2\,,\quad
D_3 = m_c^2 - k_3^2\,,
\nonumber\\
&&D_4 = m_c^2 - (k_1-k_2+k_3)^2\,,\quad
N = - (k_1-k_2)^2\,,
\label{Ix:I12}
\end{eqnarray}
and $c_i$ and $c_{ij}$ are coefficients depending on $d=4-2\varepsilon$
\begin{eqnarray*}
&&c_{10} = (d-1) (5 d^4 - 104 d^3 + 73 d^2 - 2116 d + 2086)\,,\\
&&c_{11} = (d-1) (2d-7) (2 d^3 - 35 d^2 + 180 d - 256)\,,\\
&&c_{12} = (d-9) (2d-5) (2d-7) (2d-9)\,,\\
&&c_{20} = 2 (d^4 - 22 d^3 + 165 d^2 - 491 d + 487)\,,\\
&&c_{21} = (d-9) (2d-7) (2d-9)\,,\\
&&c_{-1} = (d-3) (d-5) (d-7) (d-9) (2d-5) (2d-7) (2d-9)\,,\\
&&c_0 = (d-1) (d-3) (4 d^5 - 108 d^4 + 1090 d^3 - 5009 d^2 + 9838 d - 5335)\,,\\
&&c_1 = (d-1) (d-7) (2 d^5 - 46 d^4 + 384 d^3 - 1423 d^2 + 2158 d - 739)\,,\\
&&c_2 = (d-1) (d-5) (d-7) (d-9) (2d-7) (2d-9)\,.
\end{eqnarray*}

The master integrals used in Ref.~\cite{Bekavac:2009gz} are related to $I_{1,2}$ by
\begin{eqnarray}
I_{4.3} &=& (m_b m_c)^{2-3\varepsilon} \Gamma^3(\varepsilon) I_1(x)\,,
\nonumber\\
I_{4.3a} &=& (m_b m_c)^{1-3\varepsilon} \Gamma^3(\varepsilon) \frac{x}{1-x^2}
\nonumber\\
&&{}\times\left[ - \frac{1}{4} \left(d-3 - (2d-5) x^2\right) I_1(x)
+ \frac{3}{16} (d-2) x I_2(x)
+ \frac{x^\varepsilon + x^{2-\varepsilon}}{(d-2)^2} \right]\,.
\label{Ix:4.3}
\end{eqnarray}
Using their expansions in $\varepsilon$~\cite{Bekavac:2009gz} we obtain
\begin{equation}
I(x) = - \frac{32}{27} \left[ 1 - \frac{2}{3} \varepsilon
+ \frac{1}{2} \left( \frac{25}{3} + 3 L^2 \right) \varepsilon^2
+ B \varepsilon^3 + \cdots \right]\,,
\label{Ix:e}
\end{equation}
where
\begin{eqnarray}
&&\frac{32}{3} B = 64 L_+(x)
- \frac{(1+x^2)(5+22x^2+5x^4)}{x^3} L_-(x)
\nonumber\\
&&{} + \frac{5+18x^2+5x^4}{x^2} L^2
- 10 \frac{1-x^4}{x^2} L
+ 10 \frac{(1-x^2)^2}{x^2} + \frac{64}{3} \zeta_3 - \frac{1256}{81}\,,
\label{Ix:B}
\end{eqnarray}
and
\begin{eqnarray}
L_\pm(x) &=& L_\pm(x^{-1}) =
\Li3(x) - L \Li2(x) - \frac{L^2}{2} \log(1-x) + \frac{L^3}{12}
\nonumber\\
&&{} \pm \left[ \Li3(-x) - L \Li2(-x) - \frac{L^2}{2} \log(1+x) + \frac{L^3}{12} \right]\,,
\label{Ix:L}
\end{eqnarray}
with $L=\log x$.
Note that the functions $L_\pm(x)$ are analytical from 0 to $+\infty$.

For $x=1$, $I_2(1)$ is not independent~\cite{Broadhurst:1991fi}:
\begin{equation}
I_2(1) = - \frac{4}{3} \left( I_1(1) + \frac{8}{(d-2)^3} \right)\,.
\label{Ix:1}
\end{equation}
The expansion of $I_1(1)$ in $\varepsilon$ has been studied
in Refs.~\cite{Broadhurst:1991fi,Broadhurst:1996az}.
Using the explicit formulas~(3.2) and~(2.3) from~\cite{Bekavac:2009gz},
it is easy to get
\begin{equation}
I(1) = - \frac{32}{27} \left[ 1 - \frac{2}{3} \varepsilon + \frac{25}{6} \varepsilon^2
- \left( 7 \zeta_3 + \frac{157}{108} \right) \varepsilon^3 + \cdots \right]\,,
\label{Ix:1a}
\end{equation}
in agreement with~(\ref{Ix:e}).

For $x\to0$, two regions~\cite{Smirnov:2002pj} contribute to 
$I(x)$ (see Eq.~(\ref{Calc:Idef})), the hard ($k\sim m_b$) and 
and the soft ($k\sim m_c$) one. The
result for the leading term is given by
\begin{eqnarray}
I(x) &=& I_h x^{3\varepsilon} \left[1 + \mathcal{O}(x^2)\right]
+ I_s x^{-\varepsilon} \left[1 + \mathcal{O}(x^2)\right]\,,
\label{Ix:0}\\
I_h &=&  \frac{8}{3}
\frac{d-5}{(d-1) (d-3) (2d-9) (2d-11)}
\frac{\Gamma(1-\varepsilon) \Gamma^2(1+2\varepsilon) \Gamma(1+3\varepsilon)}%
{\Gamma^2(1+\varepsilon) \Gamma(1+4\varepsilon)}\,,
\nonumber\\
I_s &=&  \frac{8}{3}
\frac{d-6}{(d-2) (d-5) (d-7)}\,.
\nonumber
\end{eqnarray}
Expanding this formula in $\varepsilon$ we reproduce Eq.~(\ref{Ix:e}) for $x\to0$.


\section{Ghost--gluon vertex at two loops}
\label{S:Ghost}

We need this vertex expanded in the external momenta up to the linear terms.
Let us consider the right-most vertex on the ghost line:
\[
\raisebox{-7mm}{\begin{picture}(27,18)
\put(13.5,9.5){\makebox(0,0){\includegraphics{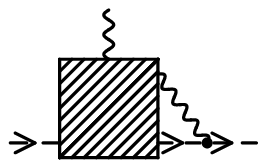}}}
\put(23.5,0){\makebox(0,0)[b]{$p$}}
\put(12.5,18){\makebox(0,0)[tl]{$\mu$}}
\put(22,4.5){\makebox(0,0)[bl]{$\nu$}}
\end{picture}}
= A^{\mu\nu} p_\nu\,.
\]
The tensor $A^{\mu\nu}$ may be calculated at zero external momenta,
hence $A^{\mu\nu}=Ag^{\mu\nu}$.
Therefore all loop diagrams have the Lorentz structure of the tree vertex,
as expected.

Now let us consider the left-most vertex:
\[
\raisebox{-7mm}{\begin{picture}(27,18)
\put(13.5,9.5){\makebox(0,0){\includegraphics{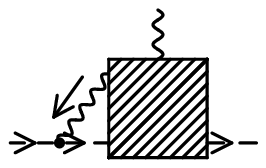}}}
\put(3.5,0){\makebox(0,0)[b]{$0$}}
\put(8.5,0){\makebox(0,0)[b]{$k$}}
\put(5,10){\makebox(0,0){$k$}}
\end{picture}}\,.
\]
It gives $k^\lambda$,
thus singling out the longitudinal part of the gluon propagator.
Therefore, all loop corrections vanish in Landau gauge.
Furthermore,
diagrams with self-energy insertions into the left-most gluon propagator
vanish in any covariant gauge: 
\[
\raisebox{-9mm}{\includegraphics{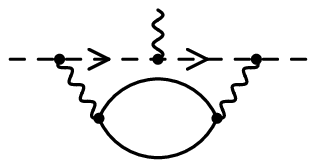}} =
\raisebox{-7mm}{\includegraphics{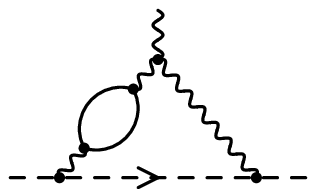}} = 0\,.
\]

In the diagrams including a quark triangle, the contraction of
$k^\lambda$ transfers
the gluon propagator to a spin 0 propagator and a factor $k^\rho$ which
contracts the quark-gluon vertex. After decomposing $k\!\!\!/$\,\, into a
difference of the involved fermion denominators one obtains in
graphical form
\begin{eqnarray*}
&&\raisebox{-7mm}{\includegraphics{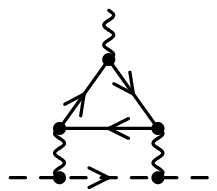}}
= a_0 \left[ \raisebox{-7mm}{\includegraphics{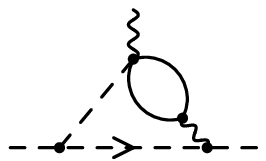}}
- \raisebox{-7mm}{\includegraphics{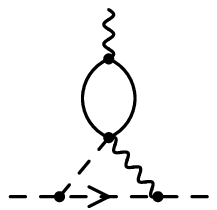}} \right]\,,\\
&&\raisebox{-7mm}{\includegraphics{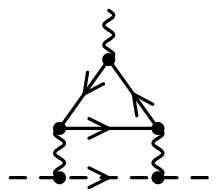}}
= a_0 \left[ \raisebox{-7mm}{\includegraphics{figs/ghb.eps}}
- \raisebox{-7mm}{\includegraphics{figs/gha.eps}} \right]\,.
\end{eqnarray*}
The diagrams with a massless triangle vanish.
The non-vanishing diagrams contain the same Feynman integral,
but differ by the order of the colour matrices along the quark line,
thus leading to a commutator of two Gell-Mann matrices.

The remaining diagram contains a three-gluon vertex with a self energy
inserted in the right-most gluon propagator. The contraction of $k^\lambda$ 
with the three-gluon vertex cancels the gluon propagator to the right of the
three-gluon vertex:
\[
\raisebox{-7mm}{\includegraphics{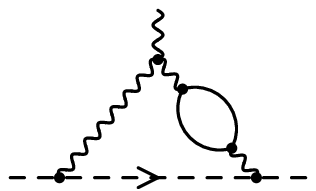}}
= a_0 \raisebox{-7mm}{\includegraphics{figs/gha.eps}}\,.
\]
The colour structure of the three-gluon vertex is identical to the
commutator above, however with opposite sign.
Therefore, after summing all contributions the result is zero.

\section{Decoupling at on-shell masses}
\label{S:OS}

For some applications it is convenient to parametrize the decoupling constants
in terms of the on-shell instead of $\overline{\mbox{MS}}$ quark masses.
The corresponding counterterm relation reads
\begin{equation}
m_{b0} = Z_{m_b}^{{\rm os}(n_f)}\left(\alpha_{s0}^{(n_f)}\right) M_b\,,\quad
m_{c0} = Z_{m_c}^{{\rm os}(n_f)}\left(\alpha_{s0}^{(n_f)}\right) M_c\,,
\label{OS:mass}
\end{equation}
where in our application $Z_{m_b}^{{\rm os}(n_f)}$ and 
$Z_{m_c}^{{\rm os}(n_f)}$ are needed to two-loop accuracy.
They have been calculated in Ref.~\cite{Gray:1990yh}
(see also~\cite{Davydychev:1998si,Bekavac:2007tk}).
Note that the two-loop coefficients of $Z_{m_b}^{{\rm os}(n_f)}$ and
$Z_{m_c}^{{\rm os}(n_f)}$ are non-trivial functions of $m_c/m_b$; a compact
expression can be found in Ref.~\cite{Bekavac:2007tk}.

The advantage of using on-shell masses is that they are identical in all
theories (with any number of flavours). Furthermore their numerical value does
not depend on the renormalization scale.
However, it is well known that usually the coefficients of perturbative series
for physical quantities grow fast when expressed via on-shell quark
masses and hence the ambiguities of the mass values
(extracted from those observable quantities) are quite large.
Nevertheless, using on-shell masses in intermediate theoretical formulae
(at any finite order of perturbation theory) can be convenient.

The decoupling relations are particularly compact if 
$\alpha_s^{(n_l)}(M_c)$ is expressed 
as a series in $\alpha_s^{(n_f)}(M_b)$ since then the
coefficients only depend on $x_{\rm os} = M_c/M_b$ (see results in~\cite{progdata}).



\end{document}